\documentclass[12pt]{report}
\usepackage[LGR,LGR,T1]{fontenc}
\usepackage{amsmath,amssymb,array}
\usepackage{amssymb}
\usepackage{textcomp}
\usepackage{amstext}
\usepackage{graphicx}
\usepackage{setspace}
\usepackage{esint}
\usepackage{caption}
\usepackage[]{latexsym}
\usepackage{epsfig}
\usepackage{amssymb}
\usepackage{makeidx}
\usepackage{longtable} 
\usepackage{eucal}
\usepackage{subfigure}
\usepackage{latexsym}
\usepackage{fancyhdr}
\usepackage{amsfonts}
\usepackage{floatflt}
\usepackage[LGR,LGR,T1]{fontenc}
\usepackage[english]{babel}
\usepackage[latin1]{inputenc}
\usepackage{bbold}
\usepackage{mathrsfs}
\usepackage{sectsty}
\usepackage{lipsum}
\usepackage{geometry}                		                  	    		
\usepackage{amscd}
\usepackage{dsfont}
\usepackage{amsthm}
\theoremstyle{plain}
\newtheorem{thm}{Theorem}[chapter]

\newtheorem{prop}[thm]{Proposition} 
\newtheorem{defn}{Definition}[chapter] 

\input xy
\xyoption{all}

\begin{document}

{\thispagestyle{empty}
\vskip 1.5cm {\Large\centerline {\bf Universit\`a degli studi di Napoli ``Federico II''}}
\vskip 1.2cm {\Large\centerline {Scuola Politecnica e delle Scienze di Base}}
\vskip 0.2cm {\Large\centerline {Area Didattica di Scienze Matematiche Fisiche e Naturali}}  
\vskip 1.5cm {\Large\centerline {\bf Dipartimento di Fisica}}
\vskip 1.5cm
\begin{center}
\includegraphics[scale=0.5]{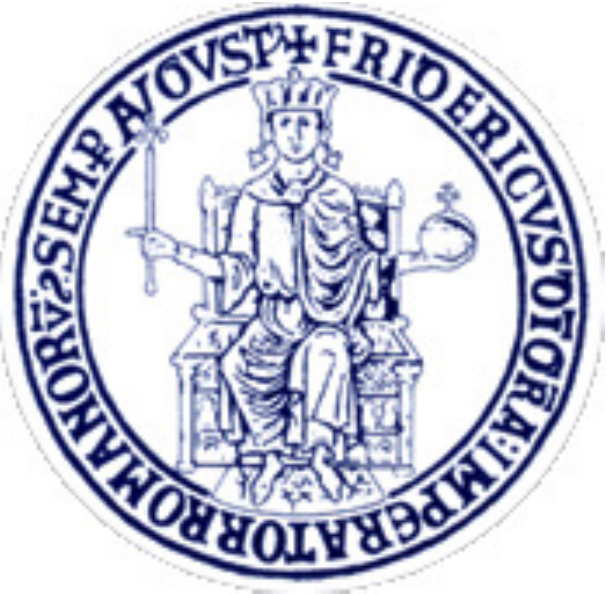}
\end{center}
\vskip 0.7cm {\Large\centerline { Laurea Magistrale in Fisica}} 
\vskip 0.7cm \centerline {Anno Accademico $2015/2016$}
\vskip 1.0cm \center{\bf{\LARGE  Interaction from Geometry,\\ Classical and Quantum}} 
\vskip 2.0cm

\begin{minipage}[t]{6cm}
Relatore

Prof. Giuseppe Marmo

\vspace{0.2cm}

%
\end{minipage}
\hfill
\begin{minipage}[t]{5cm}
\hfill Candidato

\hfill Marco~Laudato

\hfill matr. N94/249
\end{minipage}

\vfill
\eject}

\clearpage

\newpage
\null\vspace{\stretch{1}}
\begin{flushright}
\textit{Ad Elisabetta,\\
compagna di vita.}
\end{flushright}
\null\vspace{\stretch{2}}

\tableofcontents


\chapter*{Introduction}
\addcontentsline{toc}{chapter}{Introduction}
The main motivation of this work is the description of classical and quantum dynamical systems in interaction by means of geometric reduction procedures.

It is possible, indeed, to obtain non-linear, interacting, dynamical systems starting from linear dynamics such as geodesic or harmonic motions and proceeding with reduction procedures.

This work has to be considered as a first step toward the wider (and non accomplishable in a master thesis) target of the description, by means of these techiniques, of interacting quantum field theories. Therefore it is natural that we need to deal with two important extensions of particle dynamics.

The first is the passage to an arbitrary number of dimensions. In this sense, the intrinsic language of the modern Differential Geometry becomes an useful tool because it avoid us to deal with a particular system of coordinates, which may lead to cumbersome and not very fruitful computations, giving us the possibility to handle high dimensional systems without losing the general picture.

The second is the introduction in this picture of relativistic covariance. This step is needed because fundamental interactions are described by means of fields which are defined on spacetime. Therefore we will proceed to an accurate analysis of the structure of spacetime itself in terms of the language of Differential Geometry.

In this work we make a first step toward field theories by analysing the genealization of this paradigm to many-bodies dynamical system. Indeed, the investigation of aspects of field theories by means of many-bodies approximations is a well established approach in Physics.

The organization of the work is organized as follows.
In the first Chapter we will motivate the use of reduction procedures as tools to generate interacting system out of free ones by means of several elementary examples. Furthermore, we will also give a formal definition of reduction procedure. In the second Chapter, instead, we will proceed to an analysis of spacetime devoted to tensorialize the concept of reference frame, we will study in great detail the free relativistic particle, we will generalize to a many-bodies system of relativistic particles and then we will analyse the consequences of the No-Interaction Theorem.

A number of Appendices are added to elucidate some claims made in the main text.

\chapter{Geometric Reduction Procedures}
Reduction procedures, to the best of our knowledge, were introduced in a systematic way, not dealing only with specific examples, for the first time by Sophus Lie in 1893 \cite{1}. In Physics, they are used as an aid in trying to integrate dynamical systems by quadrature. Indeed, it is well known that within the Hamiltonian formalism an appropriate number of constants of the motion can be used to introduce action-angle variables which are used to analyse completely integrable systems \cite{2}\cite{3}.

From our point of view, however, we are interested in another interesting feature of reduction procedures. Indeed, it results that many non-linear systems can be related to linear ones in such a way that (obvious) integrability of the latter will entail integrability of the former. This is done by means of reduction procedures.

The aim of the first section is to motivate this proposition by means of a collection of elementary examples which are ubiquotous in many branches of Physics. In particular, we will show how to obtain very interesting dynamical systems in interaction (e.g., Riccati evolution, Calogero-Moser systems, etc.) starting from free dynamics.

The last section of this Chapter is devoted to the formal definition of reduction procedure.

\section{Examples of Reduction from Particle Dynamics}
In this Section we will sketch, by means of elementary examples, two different way of constructing nonlinear systems which are associated with a linear one and that, in addition, are integrable in the same sense as the original system is, namely:

\begin{itemize}
\item Restriction to invariant surfaces;
\item Reduction and quotienting by equivalence relations.
\end{itemize}

\noindent
To motivate our proposition that reduction procedures are able to describe interacting system out of free ones, without losing ourself in cumbersome technical details, we have chosen to present this collection of examples by using just elementary notions from calculus and the elementary theory of differential equations. The formal definition of reduction procedures will be outlined in the following Section.

\subsection{Free System and Invariant Surfaces}
In this subsection we will show, whitout any claim of generality, some examples in which we obtain interacting systems out of the free motion on $\mathbb R^3$. The main idea is to restrict Cauchy data to an invariant surface, say $\Sigma$, which is not a linear subspace of $\mathbb R^3$. It means that if $x_0$ and $x_1$ are Cauchy data on $\Sigma$, it will be not true in general that $x_0+x_1\in\Sigma$.

Let us start \cite{4}\cite{48} from Newton equations describing the free motion in $\mathbb R^3$ of a particle of unit mass:

\begin{equation}
\label{Rex1}
\ddot{\vec r}=0
\end{equation}

\noindent
which can be written in the form of a system:

\begin{equation}
\label{Rex2}
\begin{cases}
\dot{\vec r}=\vec v\\
\dot{\vec v}=0
\end{cases}
\end{equation}

\noindent
and is therefore associated to the second-order vector field $\Gamma$ in $T\mathbb R^3$:

\begin{equation}
\label{Rex3}
\Gamma=v^i\frac{\partial}{\partial r^i}.
\end{equation}

\noindent
This equation has large symmetry group\footnote{The Lie symmetry group of linear point-transformations is of dimension 12, since it is the semi-direct product $\mathbb R^3\rtimes GL(3,\mathbb R)$. Additional symmetries are obtained if we allow also non-linear transformations and transformations which are not lifted from point-transformations.} and has the constants of motion:

\begin{equation}
\label{Rex4}
\frac{d}{dt}(\vec r\wedge\dot{\vec r})=0,\quad\frac{d}{dt}\dot{\vec r}=0.
\end{equation}

\noindent
Let us introduce spherical polar coordinates:

\begin{equation}
\label{Rex5}
\vec r=r\hat n,\qquad \hat n\cdot\hat n=1,\;r=||\vec r||\geq0,
\end{equation}

\noindent
where $\hat n=\vec r/r$ is the unit vector in the direction of $\vec r$. By taking the time derivative we find:

\begin{equation}
\label{Rex6}
\dot{\vec r}=\dot r\hat n+r\dot{\hat n },\qquad\ddot{\vec r}=\ddot r\hat n+2\dot r\hat n+r\ddot{\hat n}.
\end{equation}

\noindent
By using the identities:

\begin{equation}
\label{Rex7}
\hat n\cdot \hat n=1,\quad\hat n\cdot\dot{\hat n}=0,\quad\dot{\hat n}=-\hat n\cdot\ddot{\hat n},
\end{equation}

\noindent
we obtain that the equations of motion (\ref{Rex2}), in spherical coordinates, split to two equations, one which describes the evolution "along the radius" and another "along the sphere". The radial equation and the angular momentum are:

\begin{equation}
\label{Rex8}
\begin{split}
\ddot r&=-r\dot{\hat n}^2\\
\frac{d}{dt}(\vec r\wedge\dot{\vec r})&=\frac{d}{dt}(r^2\hat n\wedge\dot{\hat n})=0.
\end{split}
\end{equation}

\noindent
By making use of the constants of the motion, we can make the first of Eqs. (\ref{Rex8}) depending only on the variable $r$ and $\dot r$ and therefore it will define a differential equation in only one degree of freedom. Indeed, by selecting a particular value of the angular momentum, say $l$ for istance, we define an invariant submanifold $\Sigma_l$, namely:

\begin{equation}
\label{Rex11}
l^2=r^4\dot{\hat n}^2
\end{equation}

\noindent
Then, by restricting the radial equation of motion to this submanifold (i.e., by solving for $\dot{\hat n}^2$ the previous equation) we describe a family of one-dimensional dynamics with different initial data but with the same value of the angular momentum, i.e.:

\begin{equation}
\label{Rex12}
\ddot r=\frac{l^2}{r^3}.
\end{equation}

\noindent
This is the first example of reduction procedure which "generates" interaction. Indeed, we have started from the \textit{free motion} on $\mathbb R^3$ and, by constraining the radial dynamics to lie on an invariant submanifold defined by fixing a constant of the motion, we have obtained a \textit{nonlinear dynamical system in interaction which exhibits a non-linear term describing an interaction}.

Obviously, we can proceed similarly by using other constants of motion, for istance energy. In this case we define an invariant surface $\Sigma_E$ by fixing the value of the energy:

\begin{equation}
\label{Rex9}
2E=\dot{\vec r}\cdot\dot{\vec r}=\dot r^2+r^2\dot{\hat n}\cdot\dot{\hat n}\,\Longrightarrow\,\dot{\hat n}^2=\frac{1}{r^2}(2E-\dot r^2)
\end{equation}

\noindent
and the equation of motion of the radial part, restricted to the constant energy surface $\Sigma_E$, will envolve only $r$ and $\dot r$:

\begin{equation}
\label{Rex10}
\ddot r=\frac{2E}{r}-\frac{\dot r^2}{r}
\end{equation}

\noindent
where $E$ plays the role of a coupling constant. 

More generally, by means of a convex combination of energy and angular momentum, i.e. $\alpha(\vec r\wedge\dot{\vec r}^2)+(1-\alpha)\dot{\vec r}^2=k$, where $0\leq\alpha\leq1$, $k\in\mathbb R$, we select and invariant submanifold $\Sigma_k$, on which the radial equations of motion becomes:

\begin{equation}
\label{Rex13}
\ddot r=\frac{\alpha l^2+(1-\alpha)(2E-\dot r^2)r^2}{r^3}.
\end{equation}

We can also select a time-dependent surface $\Sigma_t$ by fixing the value of a time-dependent constant of the motion, for istance:

\begin{equation}
\label{Rex14}
r^2+\vec v^2t^2-2\vec r\cdot\vec v\,t=k^2
\end{equation}

\noindent
we find:

\begin{equation}
\label{Rex15}
\dot{\hat n}^2=\frac{1}{r^2}\bigl[(k^2+2\vec r\cdot \vec v-r^2)t^{-2}-\dot r^2\bigr].
\end{equation}

\noindent
If we now replace Eq. (\ref{Rex15}) in the first of Eqs. (\ref{Rex8}), we find the time-dependent equation of motion:

\begin{equation}
\label{Rex16}
\ddot r=\frac{k^2}{rt^2}+2\frac{\dot r}{t}-\frac{1}{rt^2}-\frac{\dot r^2}{r}.
\end{equation}

\noindent
which is again the equation of motion of a nonlinear system in interaction.

\subsection{Calogero-Moser System}
Calogero-Moser dynamics is a nonlinear, many-body, (super) integrable, dynamical system which describes $N$ interacting particles on a line or on a circle. It is described by the following Lagrangian function:

\begin{equation}
\label{Rex17}
\mathscr L=\frac{1}{2}\sum_n^N \dot q_n^2-g^2\sum_{m\neq n}^N(q_n-q_m)^{-2},
\end{equation}

\noindent
where $g\in\mathbb R$ is a coupling constant. We will show now how obtain, by means of a reduction procedure, the Calogero-Moser dynamical system for two particles out of free motion on $\mathbb R^3$.

Let us start from $\mathbb R^3$, parametrized in terms of symmetric $2\times 2$ matrices:

\begin{equation}
\label{Rex18}
\mathbb R^3\ni(x_1,x_2,x_3)\longmapsto X=\begin{pmatrix}
x_1&\frac{x_2}{\sqrt 2}\\
\frac{x_2}{\sqrt 2}&x_3
\end{pmatrix}
\end{equation}

\noindent
The free motion equations becomes in this notation:

\begin{equation}
\label{Rex19}
\ddot X=0.
\end{equation}

\noindent
Therefore, the matrix:

\begin{equation}
\label{Rex20}
M=[X,\dot X]
\end{equation}

\noindent
is a constant of the motion, indeed:

\begin{equation}
\label{Rex21}
\dot M=[\dot X,\dot X]+[X,\ddot X]=0,
\end{equation}

\noindent
where we have used Eq. (\ref{Rex19}). The matrix $M$ is a matrix of constants of the motion whose non-zero elements are proportional to the third component $l_3$ of the angular momentum. In fact:

\begin{equation}
\label{Rex22}
M=-(x_2\dot x_3-\dot x_2 x_3)\alpha=l_3\alpha,\quad\alpha=i\sigma_2=\begin{pmatrix}
0&1\\
-1&0
\end{pmatrix}.
\end{equation}

\noindent
We want to proceed in analogy with the previous example by introducing new coordinates for the symmetric matrix $X$ by means of the rotation group. Indeed, $X$ can be diagonalized by means of an orthogonal transformation $G$, i.e.:

\begin{equation}
\label{Rex23}
X=GQG^{-1},
\end{equation}

\noindent
where:

\begin{equation}
\label{Rex24}
Q=\begin{pmatrix}
q_1&0\\
0&q_2
\end{pmatrix},\quad G=\begin{pmatrix}
\cos{\phi}&\sin{\phi}\\
-\sin{\phi}&\cos\phi
\end{pmatrix}.
\end{equation}

\noindent
In this picture, the diagonal matrix $Q$ will plays the role of "radial coordinate" while $G$ plays the role of "angular coordinate". As in the previous example, we will restrict the radial equation (i.e., the equation of motion of the diagonal matrix $Q$) to an invariant surface obtained by fixing the value of the constant of the motion. This procedure will yelds the evolution of the eigenvalues of the matrix $X$ that, being related to the trace of the powers of $X$, will be polynomial in time.

Let us show it explicitly. Since:

\begin{equation}
\label{Rex25}
GQG^{-1}=\begin{pmatrix}
q_1\cos^2\phi+q_2\sin^2\phi&(q_2-q_1)\sin\phi\cos\phi\\
(q_2-q_1)\sin\phi\cos\phi&q_1\sin^2\phi+q_2\cos^2\phi
\end{pmatrix}
\end{equation}

\noindent
we obtain the relation:

\begin{equation}
\label{Rex26}
x_1+x_3=q_1+q_2,\quad x_2=\frac{1}{\sqrt 2}(q_2-q_1)\sin 2\phi,\quad x_1-x_3=(q_1-q_2)\cos 2\phi.
\end{equation}

\noindent
Then, by using the fact that:

\begin{equation}
\label{Rex27}
\frac{d}{dt}G^{-1}=-G^{-1}\dot G G^{-1}
\end{equation}

\noindent
we can compute:

\begin{equation}
\label{Rex28}
\begin{split}
\dot X&=\dot G QG^{-1}+G\dot Q G^{-1}-GQG^{-1}\dot GG^{-1}\\
&=G\bigl([G^{-1}\dot G,Q]\bigr)G^{-1}\\
&=G\bigl(\dot Q+\dot\phi[\alpha,Q]\bigr)G^{-1}
\end{split}
\end{equation}

\noindent
where we have used:

\begin{equation}
\label{Rex29}
G^{-1}\dot G=\dot GG^{-1}=\dot\phi\alpha.
\end{equation}

\noindent
Consequently, the constant of the motion matrix $M$ becomes:

\begin{equation}
\label{Rex30}
M=[X,\dot X]=-\dot\phi(q_2-q_1)^2\sigma_1,\;\Longrightarrow\;l_3=\dot\phi(q_2-q_1)^2,
\end{equation}

\noindent
where we have used:

\begin{equation}
\label{Rex31}
[\alpha,Q]=(q_2-q_1)\sigma_1,\quad[Q,\dot Q]=0.
\end{equation}

\noindent
By taking another time derivative of Eq. (\ref{Rex28}), we obtain the following equation of motion for the diagonal matrix:

\begin{equation}
\label{Rex32}
\ddot Q-\dot\phi^2[\alpha,[\alpha,Q]]=0,
\end{equation}

\noindent
and the constant of the motion are such that:

\begin{equation}
\label{Rex33}
\frac{d}{dt}Tr(M\alpha)=0.
\end{equation}

\noindent
As in the previous example, by fixing the value of constant of the motion we define an invariant surface $\Sigma_l$:

\begin{equation}
\label{Rex34}
\Sigma_l=\left\{\frac{1}{2}Tr\,M\alpha=\dot\phi(q_2-q_1)^2\equiv g\right\},
\end{equation}

\noindent
by solving for $\dot\phi$, the "radial" equation (\ref{Rex32}) becomes:

\begin{equation}
\label{Rex35}
\ddot Q=\frac{g^2}{(q_2-q_1)^4}[\alpha,[\alpha,Q]]
\end{equation}

\noindent
which describes a family of one-dimensional dynamics with different initial data but with the same value of $l$.

At the level of the eigenvalues of $X$, we find the equations of motion:

\begin{equation}
\label{Rex36}
\ddot q_1=-\frac{2g^2}{(q_2-q_1)^3},\quad\ddot q_2=\frac{2g^2}{(q_2-q_1)^3},
\end{equation}

\noindent
which are the Euler-Lagrange equations associated with the Lagrangian (\ref{Rex17}) of the Calogero-Moser dynamics restricted at the case $N=2$.

This example is of great interest for us because it shows that reduction procedures are able to describe also interacting systems of \textit{many bodies}. In particular, it would be possible to replace matrices with selfadjoint operators on some infinite dimensional Hilbert spaces and proceed in the same manner to obtain an infinite dimensional dynamics which may describe field theories.

\subsection{Rotationally Invariant Dynamics}
Let us now consider an extension of the previous cases in which the invariant surfaces are obtained as quotient by some equivalence relation. Suppose a dynamical system which is invariant under the action of a Lie group $\mathbb G$. It is possible to determine invariant surfaces as level sets of functions or simply as subsets of the given manifolds. These sets were called \textit{invariant relations} by Levi-Civita, to distinguish them from the level sets of constants of the motion. In order to clarify this kind of reduction procedure, we will devote the rest of this subsection to some relevant examples. 

Let us consider a dynamical system on $T\mathbb R^3$ which is invariant under the action of the group $SO(3)$. If we assume that $\Gamma$ is of the second order, its expression will be:

\begin{equation}
\label{exr3}
\Gamma=\dot{\vec r}\cdot\frac{\partial}{\partial\vec r}+\vec f(r,\dot r,\vec r\cdot\dot{\vec r})\cdot\frac{\partial}{\partial \dot{\vec r}}\,,
\end{equation}

\noindent
where for the moment we do not make any further specification on the explicit form of $\vec f$. 

What we want to do is to project the dynamics (\ref{exr3}) onto the space of the orbits of the group $SO(3)$. We can parametrize this space by using the following three invariant functions:

\begin{equation}
\label{exr4}
\xi_1=\vec r\cdot\vec r,\quad\xi_2=\dot{\vec r}\cdot\dot{\vec r},\quad\xi_3=\vec r\cdot\dot{\vec r}.
\end{equation}

\noindent
The reduced dynamics is given by $L_\Gamma\xi_i=\dot\xi_i$, $i=1,2,3$, i.e.:

\begin{equation}
\label{exr5}
\begin{split}
&\frac{d}{dt}\xi_1=2\dot{\vec r}\cdot\vec r=2\xi_3\\
&\frac{d}{dt}\xi_2=2\dot{\vec r}\cdot\vec f\\
&\frac{d}{dt}\xi_3=\xi_2+\vec r\cdot\vec f.
\end{split}
\end{equation}

\noindent
As expected by the rotational invariance of the starting dynamics, $\dot\xi_i$ can be still expressed in terms of the $\xi_i$'s.

This example shows clearly that reduction procedures may not preserve the Hamiltonian or Lagrangian description which require additional structures to be defined. Indeed, it is possible to start from a space on which it is possible to define an Hamiltonian or Lagrangian function (in our example, we start from a dynamics  of the second order on $T\mathbb R^3$) and we can reduce it to a space which cannot allow it (in our example, to the odd dimensional space $\mathbb R^3$, obtaining Eqs. (\ref{exr5})).

Let us specialize this example to the case in which the starting dynamics is the free one. In this case Eqs. (\ref{exr5}) become:

\begin{equation}
\label{exr6}
\begin{split}
&\frac{d}{dt}\xi_1=2\xi_3\\
&\frac{d}{dt}\xi_2=0\\
&\frac{d}{dt}\xi_3=\xi_2.
\end{split}
\end{equation}

\noindent
Now, $\xi_2$ is a constant of the motion. Therefore, by fixing its value, we can select an invariant surface. For istance if we fix $\xi_2=k$, on the corresponding invariant surface $\Sigma_k$, the reduced dynamics\footnote{Since the dimension of the invariant surface is even, the resulting system may be described by a Lagrangian function. By redefining $\xi_1=x$ and $2\xi_3=v$, the resulting Lagrangian function will be $\mathscr L=\frac{1}{2}v^2 -2kx$.} will be a family of differential equations, each one depending on the value of $k$:

\begin{equation}
\label{exr7}
\begin{split}
&\frac{d}{dt}\xi_1=2\xi_3\\
&\frac{d}{dt}\xi_3=k.
\end{split}
\end{equation}

\noindent
and the corresponding dynamical vector field will be:

\begin{equation}
\label{exr9}
\tilde\Gamma=v\frac{\partial}{\partial x}+2k\frac{\partial}{\partial v}.
\end{equation}

\noindent
Thus, the reduced dynamics (defined on $T\mathbb R$), obtained starting from a free particle on $\mathbb R^3$, describe a particle under the effect of a constant force, for istance a particle under the effect of a uniform gravitational field or an uniform electric field.

An alternative reduced dynamics can be obtained by fixing $\xi_2=\frac{1}{\xi_1}(\xi_3^2+l^2)$, where $l^2=\xi_1\xi_2-\xi_3^2$ is the square of the angular momentum which is obviously a conserved quantity due to the rotational invariance. In this case Eqs. (\ref{exr6}) becomes:

\begin{equation}
\label{exr10}
\begin{split}
&\frac{d}{dt}\xi_1=2\xi_3\\
&\frac{d}{dt}\xi_3=\frac{\xi_3^2+l^2}{\xi_1}
\end{split}
\end{equation}

\noindent
If we now take as variables $\xi_1=\eta^2$, the first of Eqs. (\ref{exr10}) becomes $\xi_3=\eta\dot\eta$ and the second becomes, after some algebra:

\begin{equation}
\label{exr11}
\frac{d}{dt}\dot\eta=\frac{l^2}{\eta^3}.
\end{equation}

\noindent
This equation describes again two particles in interaction in the Calogero-Moser dynamical system, in the center of mass reference frame.

\subsection{Riccati Evolution (Classical Setting)}
In this subsection we will show, in both classical and quantum setting, how one can obtain the Riccati Equation out of a reduction procedure where invariant surfaces are obtained by taking the quotient by some equivalence relation.

Jacopo Riccati (and his son) in 1720 was interested in the description of the dynamical evolution of a point on the line, given by the following non-linear equation which is nowadays called Riccati Equation:

\begin{equation}
\label{ric2}
\dot \xi=c+2b\xi-a\xi^2,
\end{equation}

\noindent
where $a,b,c\in\mathbb R$ are time dependent functions. It is possible to show \cite{4} that the solutions of this equation satisfy a non-linear superposition rule. This suggests us that some reduction procedure from a linear system was unconsciounsly carried out.

Let us show how to obtain Riccati equation by reducing the following linear dynamics on $\mathbb R^2$:

\begin{equation}
\label{ric1}
\frac{d}{dt}\begin{pmatrix}x\\y\end{pmatrix}=\begin{pmatrix}b&c\\a&-b\end{pmatrix}\begin{pmatrix}x\\y\end{pmatrix}
\end{equation}

\noindent
which is equivalent to the system:

\begin{equation}
\label{ric4}
\begin{cases}
\frac{dx}{dt}=bx+cy\\
\frac{dy}{dt}=ax-by
\end{cases}
\end{equation}

\noindent
The first order vector field of the dynamics will be:

\begin{equation}
\label{ric5}
\Gamma_A=ax\frac{\partial}{\partial y}+b\left(x\frac{\partial}{\partial x}-y\frac{\partial}{\partial y}\right)+cy\frac{\partial}{\partial x}.
\end{equation}

We are ready now to perform the reduction procedure. Indeed, since the vector field $\Gamma_A$ is linear, it commutes with the Euler vector field:

\begin{equation}
\label{ric6}
\Delta=x\frac{\partial}{\partial x}+y\frac{\partial}{\partial y},
\end{equation}

\noindent
i.e., the dilation vector field in $\mathbb R^2$. It means that the algebra of invariant functions will be such that $\Delta(f)=0$, i.e. homogeneous of degree zero functions, namely $\xi=x/y$, with $y\neq0$ (or $\zeta=y/x$, with $x\neq0$). These functions parametrize the one dimensional quotient space\footnote{We have removed the origin from $\mathbb R^2$ in order to obtain a quotient manifold which is also an Hausdorff space.} $\mathbb R^2-\{0,0\}/\Delta\approx S^1$. We can now compute the dynamics restricted to the circle $S^1$ by taking the time derivative (or more precisely, by taking the Lie derivative with respect to the dynamics) of the function $\xi$, namely:

\begin{equation}
\label{ric9}
\begin{split}
\frac{d}{dt}\xi&=\frac{\dot x y-x\dot y}{y^2}\\
&=\frac{\dot x}{y}-\frac{x\dot y}{y^2}\\
&=\frac{bx+cy}{y}-\frac{x}{y^2}(ax-by)=c+2b\xi-a\xi^2
\end{split}
\end{equation}

\noindent
which is exactly Eq. (\ref{ric2}). The same procedure can be performed with the variable $\zeta=y/x$ to obtain again a Riccati equation which then arises as the restriction of the linear dynamics (\ref{ric1}) on $\mathbb R^2$ to the circle $S^1$.

\subsection{Riccati Evolution (Quantum Setting)}
Riccati equation appears also in Quantum Mechanics when we consider the space of pure states, i.e. the space of rays in a Hilbert space. It describes the quantum evolution of a $N$-level quantum system in the ray space. Motivated by the previous example, we will show how the Riccati equation can be obtained from the Shr\"odinger's equation when we perform a reduction procedure \cite{7}.

Let us consider an $N$-level quantum system. The unitary evolution operator $U(t)\in\mathbb G\equiv\mathbb U(N)$ obeys to the Shr\"odinger equation:

\begin{equation}
\label{ric10}
i\dot U(t)=H(t)U(t),\qquad U(t_0)=\mathds 1.
\end{equation}

\noindent
The solution of this equation is given by the one parameter subgroup $\mathbb H$ generated by $iH$ (if $H$ does not depend on time) acting on $\mathbb G$ from the left. Let consider the following subgroup of the unitary group $\mathbb G$:

\begin{equation}
\label{ric11}
\mathbb H=\mathbb U(n_1)\times\mathbb U(n_2)\subset \mathbb G, \quad n_1+n_2=N.
\end{equation}

\noindent
A generic element of $\mathbb H$ will be:

\begin{equation}
\label{ric12}
\begin{pmatrix}
U_1&0\\
0&U_2
\end{pmatrix}
\end{equation}

\noindent
where $U_1$ is a $n_1\times n_1$-dimensional and $U_2$ is a $n_2\times n_2$-dimensional matrix.

We are interested in reducing the dynamics (\ref{ric10}) on $\mathbb G$ to the coset space $\mathbb G/\mathbb H$. Let us consider a generic element of $\mathbb G$:

\begin{equation}
\label{ric14}
U=\begin{pmatrix}A_0&B_0\\C_0&D_0\end{pmatrix}\begin{pmatrix}U_1&0\\0&U_2\end{pmatrix}=\begin{pmatrix}A_0U_1&B_0U_2\\C_0U_1&D_0U_2\end{pmatrix}
\end{equation}

\noindent
The time evolution of this matrix is given by Eq. (\ref{ric10}) where the Hamiltonian $H$ has the form:

\begin{equation}
\label{ric15}
H(t)=\begin{pmatrix}H_1(t)&V(t)\\V(t)^\dagger&H_2(t)\end{pmatrix}
\end{equation}

\noindent
Then, we get (for brevity, we will omit the explicit dependence on time of the Hamiltonian):

\begin{equation}
\label{ric16}
\begin{split}
&i\frac{d}{dt}\begin{pmatrix}A_0&B_0\\C_0&D_0\end{pmatrix}\begin{pmatrix}U_1&0\\0&U_2\end{pmatrix}=\begin{pmatrix}H_1&V\\V^\dagger&H_2\end{pmatrix}\begin{pmatrix}A_0&B_0\\C_0&D_0\end{pmatrix}\begin{pmatrix}U_1&0\\0&U_2\end{pmatrix}\\
&i\begin{pmatrix}\dot A_0&\dot B_0\\\dot C_0&\dot D_0\end{pmatrix}\begin{pmatrix}U_1&0\\0&U_2\end{pmatrix}+i\begin{pmatrix}A_0&B_0\\C_0&D_0\end{pmatrix}\begin{pmatrix}\dot U_1&0\\0&\dot U_2\end{pmatrix}=\begin{pmatrix}H_1&V\\V^\dagger&H_2\end{pmatrix}\begin{pmatrix}A_0&B_0\\C_0&D_0\end{pmatrix}\begin{pmatrix}U_1&0\\0&U_2\end{pmatrix}\\
&i\begin{pmatrix}\dot A_0&\dot B_0\\\dot C_0&\dot D_0\end{pmatrix}=\begin{pmatrix}H_1&V\\V^\dagger&H_2\end{pmatrix}\begin{pmatrix}A_0&B_0\\C_0&D_0\end{pmatrix}-i\begin{pmatrix}A_0&B_0\\C_0&D_0\end{pmatrix}\begin{pmatrix}\dot U_1U_1^{-1}&0\\0&\dot U_2U_2^{-1}\end{pmatrix}
\end{split}
\end{equation}

\noindent
which yelds:

\begin{equation}
\label{ric17}
\begin{split}
&i\dot A_0=H_1A_0+VC_0-iA_0\dot U_1U_1^{-1}\\
&i\dot B_0=H_1B_0+VD_0-iB_0\dot U_2U_2^{-1}\\
&i\dot C_0=V^\dagger A_0+H_2C_0-iC_0\dot U_1U_1^{-1}\\
&i\dot D_0=V^\dagger B_0+H_2D_0-iD_0\dot U_2U_2^{-1}\\
\end{split}
\end{equation}

\noindent
Following the reduction procedure, we have to find the algebra of invariant functions under the right action of $\mathbb H$. We note that any function of $Z=B_0D_0^{-1}$ (or $\tilde Z=C_0A_0^{-1}$) is invariant under the right action of $\mathbb H$, indeed\footnote{Let us remark the analogy of the variable defined in the classical setting, $\xi=x/y$.}:

\begin{equation}
\label{ric18}
\begin{split}
&C_0A_0^{-1}\Longrightarrow C_0U_1U_1^{-1}A^{-1}=C_0A_0^{-1}\\
&B_0D_0^{-1}\Longrightarrow B_0U_1U_1^{-1}D^{-1}=B_0D_0^{-1}
\end{split}
\end{equation}

\noindent
\textbf{Remark: }We are limiting our procedure to the subset of $\mathbb G$ in which $A$ and $D$ are both non-singular matrices. 
\\

\noindent
Now, by using Eq. (\ref{ric17}) and the definition of $Z=B_0D_0^{-1}$, we can obtain the dynamics reduced to the quotient space $\mathbb G/\mathbb H$ by computing:

\begin{equation}
\label{ric19}
\begin{split}
i\dot Z&=i\dot B_0 D_0^{-1}-iB_0 D_0^{-1}\dot D_0D_0^{-1}\\
&=H_1B_0D_0^{-1}+V-iB_0\dot U_2U_2^{-1}D_0^{-1}-B_0D_0^{-1}V^\dagger B_0D_0^{-1}+\\
&\quad-B_0D_0^{-1}H_2-iB_0\dot U_1U_1^{-1}D_0^{-1}\\
&=H_1Z+V-iB_0\dot U_2U_2^{-1}D_0^{-1}+\\
&\quad-Z(V^\dagger B_0D_0^{-1}+H_2+iD_0\dot U_1U_1^{-1}D_0^{-1})\\
&=V+H_1Z-ZH_2-ZV^\dagger Z
\end{split}
\end{equation}

\noindent
which yelds a matrix Riccati evolution on the coset (projective) space.

\section{Generalized Reduction Procedure}
Since the main aim of this work is to make a step toward the description of fields in terms of geometric reduction procedures, we shall consider dynamical systems with a large number of degree of freedom. Therefore, keeping in mind the main features of reduction procedures outlined by means of the several elementary examples in the previous section, we will formalize these procedures in the language of differential geometry. Indeed, it allows us to deal with systems with a large number of degree of freedom without considering a particular system of coordinates.

\subsection{Reduction Procedures in Geometrical Framework}
Let consider a carrier space $M$ for the dynamical system represented by a vector field $\Gamma\in\mathfrak X(M)$. We suppose that its flow give raise to a one-parameter group of transformations:

\begin{equation}
\label{grp4}
\Phi_t\,:\,\mathbb R\times M\longrightarrow M.
\end{equation}

\noindent
In a general reduction procedure we may identify two main aspects:

\begin{itemize}

\item[i)]We consider a submanifold $\Sigma\subset M$, invariant under the evolution, i.e.:

\begin{equation}
\label{grp5}
\Phi_t(\mathbb R\times \Sigma)\subset\Sigma\quad\text{or}\quad\Gamma(m)\in T_m\Sigma,\quad\forall m\in\Sigma,\,\forall t\in\mathbb R.
\end{equation}

\item[ii)]We search for an invariant equivalence relation $\sim$ on subsets of $\Sigma$ which is compatible with $\Gamma$, i.e.:

\begin{equation}
\label{grp6}
m\sim m'\quad\Longleftrightarrow\quad\Phi_t(m)\sim\Phi_t(m'),\quad\forall m,\,m'\in\Sigma.
\end{equation}

\end{itemize}

\noindent
The \textit{reduced carrier space} is the quotient manifold $\tilde\Sigma=\Sigma/\sim$ and the \textit{reduced dynamical system} $\tilde\Gamma$ will be the projection of $\Gamma$ on $\tilde\Sigma$ along the natural projection $\pi_\sim\,:\,\Sigma\rightarrow\tilde\Sigma$.

\begin{figure}[t!]
\centering
\includegraphics[scale=0.8]{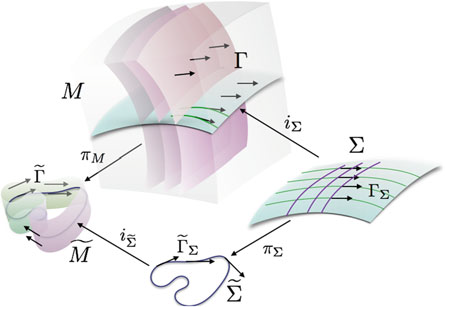}
\caption{\textit{The geometrical reduction is sketched. The purple leaves represent the $\Gamma$-invariant equivalence relation while the green surfaces is the $\Gamma$-invariant submanifold $\Sigma$. The equivalent procedure in the opposite order is also outlined (Credits: \cite{4}).}}
\label{geomred}
\end{figure}

It is possible to proceed in the opposite order, firstly considering and invariant equivalence relation $\sim$ on the whole manifold $M$ and then selecting an invariant submanifold $\tilde\Sigma\subset M$ for the reduced dynamics $\tilde\Gamma$ on the set of equivalence classes $\tilde M=M/\sim$. The geometrical reduction procedure is outlined in Fig. \ref{geomred}.

We can summarize the geometrical reduction procedure in the following \cite{4}:

\begin{thm}
\label{tgrp1}
\textbf{(Geometrical Reduction Procedure): } Let $\Gamma$ be a dynamical system defined on the carrier manifold $M$. Let $\Sigma$ be a $\Gamma$-invariant submanifold and $\sim$ a $\Gamma$-invariant equivalence relation. Let us assume that the maps $M\overset{\pi_M}{\longrightarrow}\tilde M$ and $\Sigma\overset{\pi_\Sigma}{\longrightarrow}\tilde\Sigma$, where $\tilde M=M/\sim$ and $\tilde\Sigma=\Sigma/\sim$, are smooth submersions. We denote by $\Gamma_\Sigma$ the restriction of the dynamics $\Gamma$ to $\Sigma$. The dynamical vector field $\Gamma$ is $\pi_M$-projectable and its projection to $\tilde\Sigma$ will be denoted by $\tilde\Gamma_\Sigma$. Then $\tilde\Sigma$ is a $\tilde\Gamma$-invariant submanifold in $\tilde M$ and the restriction of $\tilde\Gamma$ to it coincides with $\tilde\Gamma_\Sigma$, namely:

\begin{equation}
\label{grp7}
\tilde\Gamma_\Sigma=\tilde\Gamma\bigr|_{\tilde\Sigma}.
\end{equation}
\end{thm}

\noindent
\textbf{Remark: }Theorem (\ref{tgrp1}) cannot provide us with general conditions which ensure that the reduced carrier space is a smooth manifold. This property has to be checked on each particular instance.

\subsection{Reduction Procedures in Algebraic Framework}
It is possible to dualize the previous description in terms of the associative and commutative algebra of functions on $M$, $\mathcal F(M)$. 

We first notice that to any submanifold $\Sigma\subset M$ embedded in $M$ by means of an identification map $i_\Sigma\,:\,\Sigma\hookrightarrow M$, we can associate an associative and commutative algebra of functions on $\Sigma$, $\mathcal F(\Sigma)$, defined by $\mathcal F(\Sigma)=\mathcal F(M)/\mathcal I_\Sigma$ where $\mathcal I_\Sigma$ is a bilateral ideal defined by $\mathcal I_\Sigma=\text{Ker }i^*_\Sigma$. In other words, to any submanifold $\Sigma$ we can associate an exact sequence of associative algebras:

\begin{equation}
\xymatrix{
0\ar[r]&\mathcal I_\Sigma\ar[r]&\mathcal F(M)\ar[r]^{i^*_\Sigma}&\mathcal F(\Sigma)\ar[r]&0.
}
\end{equation}

\noindent
Thus, the two-steps procedure outlined in the previous subsection will be:

\begin{itemize}

\item[i)]The invariance condition on $\Sigma$ is obtained considering the algebra $\mathcal F(\Sigma)$ defined above and a projection $\pi_\Sigma\,:\,\mathcal F(M)\rightarrow\mathcal F(\Sigma)$ defined by:

\begin{equation}
\label{grp8}
\pi_\Sigma(f)=i^*_\Sigma(f),\quad f\in\mathcal F(M).
\end{equation}

\noindent
In this algebraic framework a dynamical system is seen as an element $\Gamma\in Der\,\mathcal F(M)$ giving:

\begin{equation}
\label{grp9}
\dot f=\Gamma\cdot f,\quad\forall f\in\mathcal F(M).
\end{equation}

\noindent
We then consider a derivation $\Gamma_\Sigma\in Der\,\mathcal F(\Sigma)$ such that:

\begin{equation}
\label{grp10}
\pi_\Sigma(\Gamma\cdot f)=\Gamma_\Sigma\cdot \pi_\Sigma(f),\quad\forall f\in\mathcal F(M).
\end{equation}

\item[ii)] In order to translate the compatibility condition we need an invariant subalgebra $\tilde{\mathcal{F}}\subset\mathcal F(\Sigma)$ such that:

\begin{equation}
\label{grp11}
\Gamma_\Sigma\cdot\tilde{\mathcal{F}}\subset\tilde{\mathcal{F}}.
\end{equation}

\noindent
The restriction of $\Gamma_\Sigma$ to $\tilde{\mathcal{F}}$ can be denoted by $\tilde\Gamma$ and represents the reduced dynamics on the $\tilde\Sigma$ of the previous subsection. We should point out that the pulled-back functions need not coincide with the algebra of all functions defined on the manifold $\Sigma$, for this to be the case we need some regularity conditions on the submanifold.

\end{itemize}


\chapter{Covariant Description of Particle Dynamics}\label{capitolodue}
As we have seen in the previous Chapter, reduction procedures allow to obtain interacting systems out of linear ones. Since the aim of this work is to make a step toward the description in these terms of field theories for fundamental interactions, it is natural to worry about relativistic covariance. Indeed, it is well known that fundamental interactions are described by means of fields defined on spacetime. Moreover, since we would deal with systems with a large number of degree of freedom, it is preferable to use the intrinsic language of modern Differential Geometry.

Therefore, in this Chapter we will analyse the structure of spacetime itself to give a definition of reference frame in intrinsic terms\footnote{In Appendix \ref{altraappendice} we will discuss also their behaviour under the action of the Poincar\'e group.}. As an illustration, we shall discuss the simple example of free relativistic particle. This will be exploited to generalize to the case of many relativistic particles in interactions and we will show a way to evade the No-Interaction Theorem and its consequences.


\section{Tensorial Characterization of Reference Frames}\label{covrefframe}

Let us start from one of the most important historical event of modern Physics. In 1905, Albert Einstein published his theory of Electrodynamics of moving bodies \cite{40} which was accepted in the body of physical science under the name of \textit{special theory of relativity}. From this moment on, the concepts of space and time assumed a completely new meaning with great consequences on the further development of Physics. Einstein arrived at this new formulation through a deep epistemological analysis of electromagnetic phenomena. The major merit of Einstein was the realization that the bearing of the Lorentz transformations transcended its connection with Maxwell's equations and was concerned with the nature of space and time in general. 

Another important contribution in the direction of a deeper understanding of the nature of space and time is the identification, due to Minkowski \cite{41}, of the carrier manifold of the perception of external word with space-time, or, in the words of Minkowski:

\begin{quotation}\noindent
"\textit{Henceforth space by itself, and time by itself, are doomed to fade away into mere shadows, and only a kind of union of the two will preserve an independent reality".}
\end{quotation}

\noindent
Einstein himself, in his book "\textit{The Meaning of Relativity}", states clearly that each "individual" should have a distinct notion of space and of time:

\begin{quote}\noindent
\textit{"The theory of relativity is intimately connected with the theory of space and time.[...] The experiences of an individual appear to us arranged in a series of events; in this series the single events which we remember appear to be ordered according to the criterion of "earlier" and "later", which cannot be analyzed further. There exists, therefore, for the individual, a subjective time. [...] it turns out that certain sense perceptions of different individuals correspond to each other, while for other sense perceptions no such correspondence can be established. We are accustomed to regard as real those sense of perceptions which are common to different individuals, and which therefore are, in a measure, impersonal. [...] We cannot speak of space in the abstract, but only of the "space belonging to a body A". [...] We shall speak only of "bodies of reference", or "space of reference".}
\end{quote}

\noindent
We shall identify the "individual" (or the "observer") with the mathematical notion of reference frame which, as emerges from the previous citation, will perform a "splitting" of space-time into space and time. The notion of \textit{mutual objective existence} will be related to a subsequent definition of compatible reference frames. In particular we will regard as real those perceptions which are common to different individuals.

What we will show in the rest of the section is the covariant formalization of the concept of reference frames. Following \cite{30, 31, 32} we will start with Weyl's analysis of space, time and matter \cite{42}:

\begin{quote}
\textit{"Time is the primitive form of the stream of consciousness. It is a fact, however obscure and perplexing to our minds, that the contents of consciousness do not present themselves simply as being, but as being now filling the form of the enduring present with a varying content. So that one does not say this is but this is now, yet now no more. If we project ourselves outside the stream of consciousness and represent its contents as an object, it becomes an event happening in time, the separate stages of which stand to one another in the relation of earlier and later. Just as time is the form of the stream of consciousness, so one may justifiably assert that space is the form of external material reality."}
\end{quote}\noindent

Existence is therefore perceived as \textit{here and now}. The mathematical object which allows this splitting of space-time into \textit{space} and \textit{time} is called a \textit{reference frame}, i.e. a rank $(1,1)$ tensor field $R$ defined on the four dimensional continuum $M$ with the property:

\begin{equation}
\label{rf1}
R\circ R=R,\quad(\text{or equivalently, } Tr\,R=1).
\end{equation}

\noindent
The fact that $R$ is a tensor, ensures that the previous definition does not depend on any specific system of coordinates. It is possible to write $R$ as the tensor product between a form $\theta\in\Omega^1(M)$ and a vector field $\Gamma\in\Lambda^1(M)$, i.e.:

\begin{equation}
\label{rf2}
R=\theta\otimes\Gamma,\quad\theta(\Gamma)=1.
\end{equation}

\noindent
Time evolution is determined by one-dimensional line transverse to the spatial leaves which are the integral curves of $\Gamma$\footnote{i.e., the well-known \textit{world lines}.}. Let us consider the tangent space $T_m M$ at a point $m\in M$. The reference frame induces a split of $T_m M$ in terms of the egeinspaces of $R(m)$ belonging to the zero eigenvalue and to the eigenvalue one, respectively:

\begin{equation}
\label{rf3}
T_m M=\text{Ker}\bigl(R(m)\bigr)\oplus\text{Im}\bigl(R(m)\bigr).
\end{equation}

\noindent
In terms of the expression (\ref{rf2}), we have $\text{Ker }R\equiv\text{Ker }\theta$ and it defines a distribution of vector fields. Due to the Frobenius' theorem, $R$ induces a foliation of space-time iff the distribution associated to the kernel of the one-form $\theta$ is involutive. Namely:

\begin{equation}
\label{rf4}
[X,Y]\in\text{Ker }\theta,\quad\forall X,Y\in\text{Ker }\theta
\end{equation}

\noindent
or, in dual way:

\begin{equation}
\label{rf5}
\theta\wedge d\theta=0.
\end{equation}

\noindent
In this case the reference frame produces a splitting of spacetime into time and space by means of a regular foliation which can be pictorially represented as a collection of pictures of the cosmos in given instants of time. It should be noticed that if we consider a different parametrization of the reference frame, for istance if the one-form becomes $e^{-f}\theta$ and the vector field becomes $e^f\Gamma$, this choice provides an equivalent splitting of spacetime and it does not affect the integrability condition (\ref{rf5}), i.e. if $\theta\wedge d\theta=0$ also $(e^{-f}\theta)\wedge d(e^{-f}\theta)=0$. 

\begin{figure}[h]
\centering
\includegraphics[scale = 0.4]{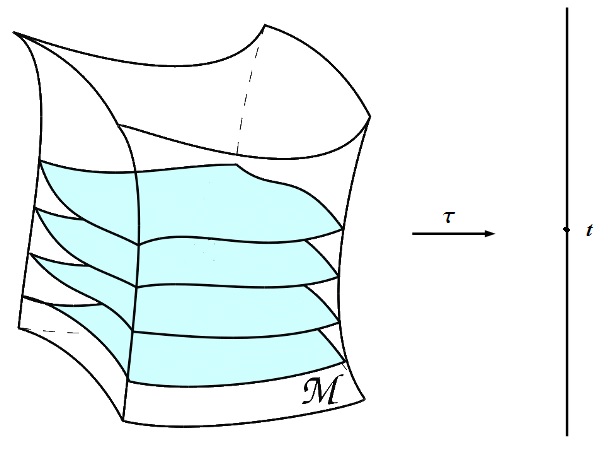} 
\caption{\textit{Simultaneity surfaces and time-axis.}}
\label{foliation}
\end{figure}

Let us give the following:

\begin{defn}
\textbf{(Synchronizable Reference Frame)}: A reference frame is said to be \textbf{synchronizable} (or to satisfy the \textbf{synchronizability condition}) if it admits a decomposition $R=\theta\otimes\Gamma$ with $\theta=d\tau$, where $\tau\in\mathcal F(M)$ is s.t.: $\tau\,:\,M\longrightarrow \mathbb R$. Level sets of the function $\tau$, $\tau^{-1}(x)\in M$, $x\in\mathbb R$ define \textbf{simultaneity leaves} (see Fig. \ref{foliation}).
\end{defn}

Finally, let us give the definition of \textit{time-like} vectors as those $v_m\in T_m M$ such that:

\begin{equation}
\label{rf6}
R(m)(v_m)=\lambda_m\Gamma(m),\quad\lambda_m\neq0.
\end{equation}

\noindent
They will be \textit{future oriented} if $\lambda_m>0$. This allows us to distinguish between past and future, indeed we can say that a point $p\in M$ is in the past of $q\in M$ if $p$ can be connected to $q$ by means of a curve whose tangent vectors are future oriented.

In conclusion, we have given a sheer covariant definition of reference frame in terms of a tensor field $R\in\mathcal T_M(1,1)$ and we have showed under which condition it induces a splitting of space-time into time and space by means of a regular foliation. We have chosen this approach because it does not depend on a metric tensor and therefore makes sense also for any Lorentzian manifold and not just Minkowski space-time\footnote{In Appendix \ref{altraappendice} we will discuss how the relativity group emerges as the group which preserves the \textit{mutual objective existence}. Moreover, we will show that the Lorentz metric tensor is \textit{derived} in this approach.}.

\section{Relativistic Free Particle}
In the previous section, we have given a tensorial characterization of the notion of reference frame. We are ready now to study a single free relativistic particle. This is a quite interesting system because, in spite of its simplicity, allows us to understand the "kinematics" of the problem, i.e. the geometric structures involved in its description.

\subsection{The Inverse Problem}
The equations of motion for a free particle\footnote{The information on the relativistic nature of the particle is not contained at the level of equations of motion because they have an infinite dimensional symmetry group. The relativity group may emerge at the level of the Lagrangian if it contains the Lorentz metric tensor.} on the Tangent Bundle $T\mathbb R^4$ are:

\begin{equation}
\label{frel1}
\begin{cases}
\frac{dx^\mu}{ds}=v^\mu\\
\frac{dv^\mu}{ds}=0
\end{cases}
\end{equation}

\noindent
where we have fixed the mass of the particle $m=1$ and $s$ is the parameter along the trajectory of the particle. The corresponding dynamical vector field is:

\begin{equation}
\label{frel2}
\Gamma=v^\mu\frac{\partial}{\partial x^\mu}.
\end{equation}

\noindent
The problem of whether or not a given dynamical vector field $\Gamma$ admits a Lagrangian description is usually known as the \textit{Inverse Problem} \cite{4}. In particular, the Partial Differential Equation that we have to solve for a Lagrangian function $\mathscr L\in\mathcal F(T\mathbb R^4)$ is:

\begin{equation}
\label{inverseproblem}
\frac{\partial^2{\mathscr{L}}}{\partial{\dot x^\mu}\partial{\dot x^\nu}}\frac{d\dot x^\nu}{ds}+\frac{\partial^2{\mathscr{L}}}{\partial{\dot x^\mu}\partial{x^\nu}}\frac{dx^\nu}{ds}-\frac{\partial{\mathscr{L}}}{\partial{x^\mu}}=0.
\end{equation}

\noindent
If the Lagrangian which solves the inverse problem is regular, i.e. its Hessian is not degenerate, $||\frac{\partial^2{\mathscr{L}}}{\partial{\dot x^\mu}\partial{\dot x^\nu}}||\neq0$, from Eq. (\ref{inverseproblem}) we can obtain the explicit form of the equations of motion (\ref{frel1}) and therefore replace the implicit differential equation associated with the Lagrangian with an explicit equation described by a vector field.

\noindent
Usually in the Literature (see for istance the Landau-Lifshitz book \cite{16}) the free relativistic particle is described by a Lagrangian function of the form:

\begin{equation}
\label{cap12.112}
\mathscr L=\sqrt{\eta_{\mu\nu}\dot x^\mu \dot x^\nu}.
\end{equation}

\noindent
This Lagrangian, defined on the full space-time and with the property that no time-parametrization is specified, i.e. it does not fix a particular splitting of space-time, solves the inverse problem (\ref{inverseproblem}) for the free motion. However, any Lagrangian function which depends only on the velocities solves it (for istance, one may consider $\mathscr L^2$). From the Lagrangian (\ref{cap12.112}), by selecting a particular time-parametrization, we can obtain other possible Lagrangians for the free relativistic particle. For istance, by defining the parameter along the trajectory of the particle as $s=x_0/c$, we obtain a Lagrangian function of the form:

\begin{equation}
\label{cap12.145}
\mathscr{L}=-c^2\sqrt{1-\frac{\vec v^2}{c^2}},\quad(\vec v=\dot{\vec r}).
\end{equation}

\noindent
In this way we fix a splitting of space-time into time and space and explicit covariance is lost\footnote{Indeed, a problem emerges with the boost. Since they mix space and time they are not compatible with a splitting of spacetime. Obviously, they are still here and emerge as nonlinear transformations which do not preserve the Tangent Bundle structure.}. It corresponds to make a gauge fixing and, obviously, it is a licit choice.

We will show now that the (gauge) freedom of time-reparametrization is allowed by the fact that the Lagrangian (\ref{cap12.112}) is not regular. Indeed, if we compute its Hessian, it results degenerate. It means that we cannot invert Eq. (\ref{inverseproblem}) and we cannot exhibit a normal form of the equations of motion. What we can do is to look for a family of vector fields which characterizes the solution of the Euler-Lagrange equation, which in intrinsic terms, are:

\begin{equation}
\label{eulerlagrange}
L_X\theta_{\mathscr L}=d\mathscr L
\end{equation}

\noindent
where $X$ is the vector field which we are looking for and $\theta_{\mathscr L}$ is the Cartan $1$-form (see Appendix \ref{s21}). For the Lagrangian (\ref{cap12.112}) it assumes the coordinate expression:

\begin{equation}
\label{cap12.113}
\theta_\mathscr{L}=\frac{\partial{\mathscr{L}}}{\partial{\dot x_\mu}}dx^\mu=\frac{1}{\mathscr{L}}(\eta_{\mu\nu}\dot x^\mu dx^\nu).
\end{equation}

\noindent
The associated Lagrangian $2$-form is:

\begin{equation}
\label{cap12.114}
\omega_\mathscr{L}=d\theta_\mathscr{L}=\frac{1}{\mathscr{L}^3}(\eta_{\mu\nu}\mathscr{L}^2-\dot x^\mu\dot x^\nu)\,d\dot x^\mu\wedge dx^\nu.
\end{equation}

\noindent 
The kernel of this 2-form is generated by:

\begin{equation}
Ker\,\omega_{\mathscr L}=\left\{\dot x^\mu\frac{\partial}{\partial x^\mu},\dot x^\mu\frac{\partial}{\partial\dot x^\mu}\right\}\equiv\left\{\Gamma,\Delta\right\}
\end{equation}

\noindent
where $\Gamma$ is the vector field (\ref{frel2}) and $\Delta$ is the Euler vector field, i.e. the vector field which generates dilation transformations along the fibers. By using the Lie identity $L_X=i_Xd+di_X$ we can recast the Euler-Lagrange equation (\ref{eulerlagrange}) in terms of the lagrangian energy:

\begin{equation}
\label{eulerlagrange2}
i_X\omega_{\mathscr L}=-d\left(\frac{\partial\mathscr L}{\partial\dot x^\mu}\dot x^\mu-\mathscr L\right)\equiv=-dE_{\mathscr L}.
\end{equation}

\noindent
But we notice that the energy function is identically zero, i.e. $E_{\mathscr L}=0$. Therefore the Euler-Lagrange equations reduce to:

\begin{equation}
i_X\omega_{\mathscr L}=0.
\end{equation}

\noindent
It means that the vector field that we are looking for is in the kernel of the Lagrangian 2-form and therefore no time parametrization is selected\footnote{It is a consequence of the fact that the constraint $E_{\mathscr L}=0$ induces a gauge freedom, in this case time reparametrization.}. Indeed, since any vector field in the kernel of $\omega_{\mathscr L}$ multiplied by an arbitrary function is still in the kernel of $\omega_{\mathscr L}$, a priori there is no reason to choice a particular vector field (which of course will carry a particular time parametrization) among the many. Only an additional requirement, such as second order dynamics, will fix univocally a particular vector field in this family. 

\subsection{Newton - Wigner Positions}
\label{subsnwp}
In the previous subsection, we have shown that, if we want to describe a free relativistic particle, we need to introduce a degenerate Lagrangian 2-form. This means that if we want to define canonical formalism\footnote{See Appendix \ref{s21} for a detailed presentation of Poisson realization of a presymplectic manifold.} on the presymplectic manifold\footnote{Parenthetically, we recall that a presymplectic manifold $(M,\omega)$ is a manifold $M$ equipped with a closed and degenerate $2$-form $\omega$.} $(T\mathbb R^4, \omega)$ we need to define the quotient space $(T\mathbb R^4)/Ker\,\omega_\mathscr L$. 

Taking the quotient by $\text{Ker }\omega_\mathscr{L}$ means that we are considering the involutive distribution generated by:

\begin{equation}
\label{cap12.131}
\Gamma=\dot x^\mu\frac{\partial}{\partial{x^\mu}},\quad\Delta=\dot x^\mu\frac{\partial}{\partial{\dot x^\mu}}.
\end{equation}

\noindent
By saying involutive we mean:

\begin{equation}
\label{cap12.132}
[\Delta,\Gamma]=\Gamma.
\end{equation}

\noindent
Starting from $\text{dim }T\mathbb R^4=8$ and removing two dimensions (because the integral leaves of the distribution (\ref{cap12.131}) are $2$-dimensional) we have:

\begin{equation}
\label{cap12.133}
\text{dim }T\mathbb R^4/\text{Ker }\omega_\mathscr{L}=6,
\end{equation}

\noindent
Since we have taken the quotient w.r.t. the dynamics and the Euler vector field, this space is parametrized by constants of the motion with the additional requirement that they are homogeneous of degree zero with respect to the velocities.

Let us now analyze the constants of the motion\footnote{It should be stressed that relation between constants of the motion and symmetries depends on a rank-2 tensor such as the Lagrangian $2$-form or the symplectic form.} associated with the equations of motion (\ref{frel1}). It is easy to see that the functions:

\begin{equation}
\label{cap12.128}
J^{\mu\nu}=\dot x^\mu x^\nu-\dot x^\nu x^\mu
\end{equation}

\noindent
are constants of the motion. Indeed, since $\ddot x^\mu=0$, we have:

\begin{equation}
\label{cap12.129}
\frac{d}{dt}(\dot x^\mu x^\nu-\dot x^\nu x^\mu)=\dot x^\mu\dot x^\nu-\dot x^\nu\dot x^\mu=0.
\end{equation}

\noindent
Eq. (\ref{cap12.128}), due to its skew-symmetry, identifies six constants of the motion which, together with the Lagrangian function (\ref{cap12.112}), form a (maximal) set of seven functionally independent constants of the motion. Since $dim\,T\mathbb R^4$=8, by fixing the value of these constants of the motion we identify a dimension one subset of $T\mathbb R^4$ which is the trajectory of the particle.


The parametrization of the quotient space $T\mathbb R^4/Ker\,\omega_{\mathscr L}$ can be accomplished, for istance, by dividing the constants of the motion (\ref{cap12.128}) by the Lagrangian, i.e.:

\begin{equation}
\label{cap12.134}
\frac{1}{\mathscr{L}}(\dot x^\mu x^\nu-\dot x^\nu x^\mu).
\end{equation}

\noindent
Indeed, $\mathscr{L}$ is homogeneous of degree one in the velocities.

Since we have taken the quotient of $T\mathbb R^4$ w.r.t. the kernel of $\omega_\mathscr L$, we can define a $2$-form $\tilde\omega$, which is not degenerate, on $T\mathbb R^4/\text{Ker }\omega_\mathscr{L}$ by:

\begin{equation}
\label{cap12.135}
\pi^*_{\text{Ker}}(\tilde\omega)=\omega_\mathscr{L},
\end{equation}

\noindent
where $\pi^*_{\text{Ker}}$ is the pull-back of the projection:

\begin{equation}
\label{cap12.136}
T\mathbb R^4\overset{\pi_{\text{Ker}}}{\longrightarrow}T\mathbb R^4/\text{Ker }\omega_\mathscr{L}.
\end{equation}

\noindent
It means that the quotient space $T\mathbb R^4/\text{Ker }\omega_\mathscr{L}$ is a symplectic manifold\footnote{
This symplectic manifold is sometimes called \textit{manifold of motions}.}. Now, being $\tilde\omega$ symplectic, \textit{Darboux's Theorem} affirms that there will be a chart, called \textit{Darboux's Chart} on $T\mathbb R^4/\text{Ker }\omega_\mathscr{L}$, such that:

\begin{equation}
\label{cap12.139}
\tilde\omega=\sum^3_{a=1}dP_a\wedge dQ_a
\end{equation}

\noindent
where the sum goes from $1$ to $3$ because $\text{dim }T\mathbb R^4/\text{Ker }\omega_\mathscr{L}=6$. As we will see in the next subsection, these coordinates $\{P,Q\}$ are "positions" and "momenta" only in the sense that they define canonical commutation relation w.r.t. the PB associated to the symplectic structure $\tilde\omega$. It results clearer if we express Darboux coordinates in terms of the variables of $T\mathbb R^4$:

\begin{equation}
\label{cap12.140}
Q^j=\mathscr{L}\frac{G^j}{\dot x^0},\quad P^j=\frac{\dot x^j}{\mathscr{L}}
\end{equation}

\noindent
where 

\begin{equation}
\label{cap12.141}
G^j=\frac{1}{\mathscr{L}}(\dot x^j x^0-\dot x^0 x^j).
\end{equation}

\noindent
If we write the first of Eqs. (\ref{cap12.140}) explicitly, we find that the \textit{Newton-Wigner positions} $Q$'s are given by:

\begin{equation}
\label{cap12.143}
Q^j=-x^j+\frac{\dot x^j}{\dot x^0}x^0
\end{equation}

\noindent
i.e., the expression of the Newton-Wigner positions $Q$  in terms of the Tangent Bundle variable contains also the physical velocities.  

\subsection{Poisson Brackets}
We are ready to define Poisson Brackets on the presymplectic manifold $T\mathbb R^4$. Following Appendix \ref{s21}, we have to define a generalized Lorentz-invariant flat connection $A_{\mathscr L}$, i.e. a $(1,1)$-tensor:
\
\begin{equation}
\label{cap12.164}
A_\mathscr{L}=\mathds 1-\left[\frac{\dot x^\mu d\dot x_\mu}{\mathscr{L}^2}\otimes\dot x^\rho\frac{\partial}{\partial{\dot x^\rho}}+\frac{1}{\mathscr{L}}d\left(\frac{\dot x^\mu x_\rho}{\mathscr{L}}\right)\otimes\dot x^\rho\frac{\partial}{\partial{x^\rho}}\right]
\end{equation}

\noindent
where we have used the fact that the identity minus a projector is still a projecton (then requirement (\ref{pre4}) is satisfied). Since the connection must be invariant with respect to the involutive distribution generated by the Kernel of $\omega_\mathscr L$, we have divided by $\mathscr L$ in order to obtain terms homogeneous of degree zero in the velocities\footnote{Lorentz invariance does not fix uniquely the form of the connection $A_{\mathscr L}$. Indeed, it is possible to multiply the vector fields in the connection by Lorentz invariant functions. Therefore, by choosing a particular connection we fix the gauge freedom on time reparametrization.}. The role of this connection is to take a vector field on the quotient manifold $T\mathbb R^4/Ker\,\omega_{\mathscr L}$ and fixing its vertical part as a vector in the kernel of $A_{\mathscr L}$. In this way, with a slight abuse of notation, it maps $A_{\mathscr L}\,:\,\mathfrak X(T\mathbb R^4/Ker\,\omega_{\mathscr L})\longrightarrow \mathfrak X(T\mathbb R^4)$.

Now we can define a PB associated to the presymplecitc structure $\omega_\mathscr L$ by computing:

\begin{equation}
\label{cap12.165}
\Lambda_\mathscr{L}=\mathscr{L}\left(A_\mathscr{L}\left(\frac{\partial}{\partial{\dot x^\mu}}\right)\wedge A_\mathscr{L}\left(\frac{\partial}{\partial{x^\mu}}\right)\right)
\end{equation}

\noindent
which is manifestly Lorentz-invariant. Here, $\mathscr{L}$ is a conformal factor and we could also remove it. In general this is not possible because by multiplying a Poisson tensor by a function it may happen that it does not satisfy the Jacobi identity anymore. However, in the case in exam, $\mathscr{L}$ is a Casimir for this PB and then it does not affect the Jacobi identity.

With this PB we have that the Darboux's coordinates are canonical (this is exactly the way we can define them), i.e.:

\begin{equation}
\label{pdb2}
\begin{split}
&\left\{Q^i,Q^j\right\}=0\\
&\left\{Q^i,P^j\right\}=\delta^{ij}\\
&\left\{P^i,P^j\right\}=0
\end{split}
\end{equation}

\noindent
while the physical coordinates (positions and velocities) do not commute anymore \cite{15}. Indeed, if we now reintroduce the constants $m$ and $c$, in the Lagrangian and in the symplectic structure:

\begin{equation}
\mathscr L=mc\sqrt{\eta_{\mu\nu}\dot x^\mu\dot x^\nu},\quad\omega_{\mathscr L}=\frac{cm}{v^3}(\eta_{\mu\nu}v^2-\dot x^\mu\dot x^\nu)\,dx^\mu\wedge d\dot x^\nu
\end{equation}

\noindent
we obtain the following Poisson Brackets:

\begin{equation}
\label{pbd3}
\begin{split}
&\left\{\dot x^\rho,x^\sigma\right\}=\mathscr L\left(g^{\rho\sigma}-m^2c^2\frac{\dot x^\rho\dot x^\sigma}{\mathscr L^2}\right)\\
&\left\{\dot x^\rho,\dot x^\sigma\right\}=0\\
&\left\{x^\rho,x^\sigma\right\}=m^2c^2\frac{\dot x^\sigma x^\rho-\dot x^\rho x^\sigma}{\mathscr L}
\end{split}
\end{equation}

\noindent
It means that the price we pay to give a covariant description in terms of Poisson Brackets is that we lose localizability of the particle. However, altough this result can embarrass, it plays a fundamental role when we want to describe the interaction between two or more relativistic particles. Indeed, as we will see in the next Section, it supplies us with a way to evade the No Interaction Theorem.

Moreover, this result can be extended to many particles and eventually to field theories, giving rise to what is known as Peierls Brackets which is a fully covariant bracket for fields and may be used for a covariant quantization of fields.

\section{Relativistic Particles in Interaction}
In this section we will investigate on the interaction between relativistic particles. The generalization w.r.t. the single particle case that we have discussed in the previous section is not trivial due to the No-Interaction Theorem. The aim of this section is to show one of the possible way to \textit{evade} this theorem.

\subsection{No-Interaction Theorem}\label{s231}
The problem of describing relativistic interacting particles in the Hamiltonian formalism goes back to Dirac \cite{23} which proposed a general framework to describe such a systems, called Generator Formalism. In this approach, one consider a phase space structure on which the Poincar\'e group is realized by canonical tranformations (i.e., transformations which preserve the symplectic structure). The fundamental idea is that one subsumes the question of dynamical evolution within the more general question of representing a change of inertial frame corresponding to any element of the Poincar\'e group. In this way, equation of motions were particular cases of more general equations describing the effect of infinitesimal Poincar\'e transformations. Thus both Hamiltonian and "interactions" are contained among the ten generators of the Poincar\'e group.

Soon after Dirac's work, several authors recognized that a fundamental ingredient to describe \textit{physical} relativistic particles in interaction is the so-called \textit{World Line Condition} (WLC). The idea here is that when the observations in two inertial frames are related by the canonical transformations representing the appropriate element of the Poincar\'e group, then in any particular state of motion, the two observers "see" the same set of world lines in spacetime. We shall refer to this requirement as WLC\footnote{We should also require the so-called \textit{separability condition} i.e. the requirement that if we move at infinity the particles from each others, each particle follows its own dynamics unaffected by the presence of the others. Since we will focus on the two-particle case, in which this condition is trivially satisfied, we will ignore this problem. However, for istance in scattering theory, this property plays a fundamental role. For further details see \cite{19}.} \cite{18} (a formal statement of WLC will be made in the following subsection).

However, if we assume that the Poincar\'e group acts as canonical transformations on the phase-space of a set of relativistic particles and that the WLC holds, a theorem, usually called \textit{No-Interaction Theorem} \cite{20, 21,22, 23}, affirms that there cannot be interactions between the particles\footnote{A proof of this theorem is given in Appendix \ref{appb}.}.

Therefore, if we want to describe a relativistic many-bodies system in interaction, we have to find a way to evade No-Interaction Theorem. Since it represents undeniably an interesting challenge for Physicists and Mathematicians, it should not surprise that  in the Literature there are several models constructed to evade this theorem. 

We will discuss a general method of dealing with all these models and then we will carry out an example.

\subsection{World Line Condition and Poincar\'e Canonical Transformation}\label{s232}
The aim of this subsection is to give a formal definition of the WLC in the Hamiltonian formalism for a free relativistic particle. It is a preparatory step to illustrate how to evade the No-Interaction Theorem. We will do it within the more general fashion of the Dirac constraints theory and reduction procedures. In particular, we will show that among the constraints that we will impose, we have to include one which explicitly depends on a parameter $\tau$, which will be identified with the evolution parameter. We have thus the curious situation in which \textit{motion is generated by constraints}.

Let us consider a single relativistic free particle. The carrier space of such a dynamical system is, in the Hamiltonian framework, $T^*\mathbb R^4$ with independent variables $(x^\mu,p^\mu)$. Since the phase-space is obviously symplectic, there are canonical PB defined on it. With respect to these brackets, the variables of the phase-space satisfy the canonical commutation relations:

\begin{equation}
\label{wlc1}
\{x^\mu,x^\nu\}=0,\quad\{x^\mu,p^\nu\}=g^{\mu\nu},\quad\{p^\mu,p^\nu,\}=0.
\end{equation}

\noindent
The Poincar\'e group acts preserving the PB structure as:

\begin{equation}
\label{wlc2}
x^\mu-x'^\mu=\Lambda^\mu_\nu x^\nu+a^\mu,\quad p^\mu-p'^\mu=\Lambda^\mu_\nu p^\nu
\end{equation}

\noindent
and has the following set of infinitesimal generators:

\begin{equation}
\label{wlc3}
J_{\mu\nu}=x_\mu p_\nu-x_\nu p_\mu,\quad P_\mu=p_\mu
\end{equation}

\noindent
which reproduce the Lie algebra of the Poincar\'e group w.r.t. the canonical PB structure. 

Let us impose the constraint:

\begin{equation}
\label{wlc4}
K=p^2-m^2=0
\end{equation}

\noindent
The constraints (\ref{wlc4}), which is the mass-shell relation for the particle, identify an hypersurface $\Sigma\subset T^*\mathbb R^4$ of codimension 1. This surface is clearly invariant under the action of the Poincar\'e group because\footnote{Parentetically, we recall that two functions $f$, $g$ on phase space are \textit{weakly equal}, $f\approx g$, if they are equal on the surface defined by the constraints and not throughout the whole phase space.}:

\begin{equation}
\label{wlc5}
\{K,J_{\mu\nu}\}\approx0,\quad\{K,P_\mu\}\approx0.
\end{equation}

\noindent
Moreover, the constraint (\ref{wlc4}) is the generator of canonical transformations mapping $\Sigma$ in itself. Indeed, starting from a point $(x,p)\in\Sigma$ we can apply to it the one parameter family of canonical transformations generated by $K$. The orbits of these transformations will be all lines $L$ which lie in $\Sigma$. In particular we can set up a system of differential equation w.r.t. an unspecified independent parameter $\tau$:

\begin{equation}
\label{wlc6}
\frac{dx^\mu(\tau)}{d\tau}\approx v\{x^\mu(\tau),K\},\quad\frac{dp^\mu(\tau)}{d\tau}\approx v\{p^\mu(\tau),K\}
\end{equation}

\noindent
where $v$ is the parameter of the family of canonical transformations. This lines foliated $\Sigma$ and the conditions (\ref{wlc5}) affirms that the Poincar\'e group maps lines in $\Sigma$ in other lines in $\Sigma$.

We have now to impose another constraint. Its purpose is to assigne at each point on a line $L$ a definite value of the evolution parameter $\tau$. To serve this purpose, the constraint must depend explicitly on $\tau$ and it must vary along $L$ for a fixed value of $\tau$. Then, the functional form of this constraint will be:

\begin{equation}
\label{wlc7}
\chi=\chi(x,p,\tau),\quad\{\chi,K\}\neq0
\end{equation}

\noindent
i.e., it is a second class constraint. When we impose both the constraints (\ref{wlc4}) and (\ref{wlc7}) we identify a six-dimensional hypersurface $\tilde\Sigma\subset T^*\mathbb R^4$. Since we have a set of second class constraints we can define the Dirac Bracket (DB) determined by $K$ and $\chi$:

\begin{equation}
\label{wlc8}
\{f,g\}_D=\{f,g\}-\bigl(\{f,K\}\{\chi,g\}-\{f,\chi\}\{K,g\}\bigr)/\{\chi,K\}.
\end{equation}

\noindent
Since the generators of the Poincar\'e group (\ref{wlc3}) have the same commutation relation also w.r.t. the DB we can realize the Poincar\'e group with transformations which are canonical w.r.t. the DB. The additional featuring which emerges is that with this new realization, the Poincar\'e group preserves the value of $\tau$ when it carries each point of a line $L$ to its image $L'$.

Having specified the evolution parameter, we can remove the arbitrariness on the parameter $v$ in Eqs. (\ref{wlc6}) by:

\begin{equation}
\label{wlc9}
\begin{split}
\frac{d\chi}{d\tau}&=\frac{\partial\chi}{\partial\tau}+\frac{\partial\chi}{\partial\chi^\mu}\frac{dx^\mu}{d\tau}+\frac{\partial\chi}{\partial p^\mu}\frac{dp^\mu}{d\tau}\\
&=\frac{\partial\chi}{\partial\tau}+\frac{\partial\chi}{\partial x^\mu}v\{x^\mu,K\}+\frac{\partial\chi}{\partial p^\mu}v\{p^\mu,K\}\\
&=\frac{\partial\chi}{\partial\tau}+v\left[\left\{\frac{\partial\chi}{\partial x^\mu}x^\mu+\frac{\partial\chi}{\partial p^\mu}p^\mu,K\right\}\right]\\
&=\frac{\partial\chi}{\partial\tau}+v\{\chi,K\}\approx0
\end{split}
\end{equation}

\noindent
and then:

\begin{equation}
\label{wlc10}
v=\frac{\partial\chi}{\partial\tau}\bigl/\{\chi,K\}.
\end{equation}

\noindent
Now we can define a new "Hamiltonian" function (in the sense that it acts as the generator of dynamical evolution via the DB) by setting:

\begin{equation}
\label{wlc11}
\frac{df}{d\tau}\approx\frac{\partial f}{\partial\tau}-\frac{\{f,K\}}{\{\chi,K\}}\approx\frac{\partial'f}{\partial\tau}+\{f,\mathcal H\}_D
\end{equation}

\noindent
where $\partial'/\partial\tau$ means the derivation w.r.t. the explicit dependence on $\tau$.

Now we are ready to discuss the WLC. Let $O$ and $O'$ be two inertial frames connected by an infinitesimal element $(\Lambda, a)$ of the Poincar\'e group. This means that the spacetime coordinates $x^\mu,x'^\mu$ are related geometrically by:

\begin{equation}
\label{wlc12}
x'^\mu=x^\mu+\omega^{\mu\nu}x_\nu+a^\mu
\end{equation}

\noindent
where $\omega^{\mu\nu}=-\omega^{\nu\mu}$ and $|\omega|,|a|\ll1$. Then we can write the generators of the Poincar\'e group as:

\begin{equation}
\label{wlc13}
G=\frac{1}{2}\omega^{\alpha\beta}J_{\alpha\beta}-a^aP_a.
\end{equation}

\noindent
We can construct the world lines in spacetime by using the canonical projection $T^*\mathbb R^4\overset{\pi}{\longrightarrow}\mathbb R^4$. Indeed, for a particle in a state of motion corresponding to the line $L\subset\tilde\Sigma\subset T^*\mathbb R^4$ we consider only the spacetime position vector $x^\mu(\tau)$. The world line is obtained by plotting the spatial positions $\vec x(\tau)$ at the laboratory time $x^0(\tau)$ (we have performed a splitting of spacetime into space and time). From the previous considerations about the realization of the Poincar\'e group w.r.t. the DB, we know that the line $L$ is carried by the infinitesimal Dirac-canonical transformation generated by $G$ into a line $L'$ preserving the value of $\tau$ by:

\begin{equation}
\label{wlc14}
x'^\mu(\tau)\approx x^\mu(\tau)+\{G,x^\mu(\tau)\}_D.
\end{equation}

\noindent
We can now state the following fundamental \cite{22}:

\begin{prop}
\textbf{(World Line Condition): }The spacetime constructions carried out in $O$ and $O'$ describe one and the same objectively real world line if, for each $\tau$, $x^\mu(\tau)$ obtained by the canonical transformation (\ref{wlc14}) is related in the geometrical manner of Eq. (\ref{wlc12}) to $x^\mu(\tau+\delta\tau)$ for some infinitesimal $\delta\tau$., i.e.:

\begin{equation}
\label{wlc15}
x'^\mu(\tau)\approx x^\mu(\tau+\delta\tau)+\omega^{\mu\nu}x_\nu(\tau+\delta\tau)+a^\mu.
\end{equation}
\end{prop}

\noindent
Here $\delta\tau$ is permitted to be a linear expression in $\omega^{\mu\nu}$ and $a^\mu$ with coefficients that could depend on dynamical variables. By considering Eqs. (\ref{wlc11}) (\ref{wlc14}) and (\ref{wlc15}) we can give the following:

\begin{defn}
\textbf{(World Line Condition): }The WLC is the condition that there exists an expression for $\delta\tau$ such that:

\begin{equation}
\label{wlc16}
\{G,x^\mu\}_D=\omega^{\mu\nu}x_\nu+a^\mu+\left(\frac{\partial'x^\mu}{\partial\tau}+\{x^\mu,\mathcal H\}_D\right)\delta\tau.
\end{equation}
\end{defn}

\noindent
In this form, the WLC is written only in terms of the Dirac Bracket and it is a condition on both $G$ and $H$.

\subsection{Evading the No-Interaction Theorem}\label{s233}

First, let us describe the general geometric picture of the models used in the Literature to evade the theorem \cite{20}. Then, we will discuss an example. 

On a given carrier space $T^*Q$ of dimension $2n$ a set of real functions $K_1,\dots,K_k$ is given. By fixing their value, they identify an hypersurface $M\subset T^*Q$ of codimension $k$. In particular, if we consider the smooth map $\varphi:T^*Q\longrightarrow\mathbb R^k$ by

\begin{equation}
\label{anit1}
\gamma\mapsto\bigl((K_1(\gamma),\dots,K_k(\gamma)\bigr),
\end{equation}

\noindent
the surface $M$ is determined by:

\begin{equation}
\label{anit2}
M=\varphi^{-1}(0)
\end{equation}

\noindent
and if $0\in\mathbb R^k$ is a regular value, then $M$ is a submanifold. By using the canonical symplectic structure $\omega$ it is possible to associate a vector field $X_j$ to the function $K_j$ by:

\begin{equation}
\label{anit3}
i_{X_j}\omega=dK_j.
\end{equation}

\noindent
It is possible to prove \cite{20} that the set of the vector fields associated with the functions $K$ satisfy the condition of the Frobenius theorem, i.e. they constitute the tangent space of a submanifold.

Let consider the identification map:

\begin{equation}
\label{anit5}
i\,:\,M\longrightarrow T^*Q.
\end{equation}

\noindent
We can consider a $2$-form on $M$ which is the pull-back of the canonical symplectic 2-form by the identification map, namely:

\begin{equation}
\label{anit6}
\omega_M=i^*\omega.
\end{equation}

\noindent
In general $\omega_M$ will be degenerate. Let us suppose that the rank of $\omega_M$ is constant. In this case the set of vector fields, say $Y_j$, which are in the kernel of $\omega_M$ constitutes an involutive distribution $\mathscr D$ (i.e., they obey at the Frobenius theorem). These vector fields are a combinations of the Hamiltonian vector fields associated to the constraints $K$. Indeed, since they are in the kernel of $\omega_M$, we have:

\begin{equation}
\label{anit7}
i_Y\omega_M=0
\end{equation}

\noindent
and then

\begin{equation}
\label{anit8}
i_Y\omega=c^i dK_i\quad\Longrightarrow\quad Y=c_i X_{K_i},
\end{equation}

\noindent
where the $c_i$ are functions on $M$. This means that:

\begin{equation}
\label{anit9}
c_i\{K_i,K_j\}\approx0
\end{equation}

\noindent
i.e., the $K$'s are first class constraints on $M$. Let us foliate $M$ by the involutive distribution $\mathscr D$, namely:

\begin{equation}
\label{anit10}
N=M/\mathscr D.
\end{equation}

\noindent
We assume $N$ to be a manifold whose dimension is $2n-k-rank\,\omega_M$. Being $N$ a manifold, the map:

\begin{equation}
\label{anit11}
\pi\,:\,M\longrightarrow N
\end{equation}

\noindent
is a submersion. Since we have taken the quotient w.r.t. the kernel of the 2-form $\omega_M$, $N$ will be a symplectic manifold. It is the analogue of the manifold of motion that we have defined in the free relativistic particle case. On this manifold, no dynamics has been defined. This is done if we assume that there exists a one-parameter family of sections:

\begin{equation}
\label{anit12}
N\times\mathbb R\overset{\sigma}{\longrightarrow} M
\end{equation}

\noindent
for the fiber bundle $M\overset{\pi}{\longrightarrow}N$. Therefore, the dynamics will be defined on $\sigma(N\times\mathbb R)\subset M$, not on $M$ itself, and the Cauchy surface will be $\sigma(N\times\{0\})$. Among all the possible dynamics that can be defined corresponding to different choices of $\sigma$, we will consider interacting system. 

We will show, in an example \cite{22}, that the No-Interaction Theorem can be evaded in this fashion because on $\sigma(N\times\mathbb R)$ the fundamental requirement for the proof the theorem, i.e. the existence of a Lagrangian function for the dynamics, does not hold\footnote{See Appendix \ref{appb} for the details.}.

Let us consider the case of two relativistic particles. The carrier space of the dynamics is $T^*\mathbb R^8$ with variables $x^\mu_\alpha$, $p^\mu_\alpha$, where $\alpha=1,2$ labels the particle. On this space it is defined a canonical symplectic structure and the non-zero commutation relation between the coordinates is:

\begin{equation}
\label{anit14}
\{x^\mu_\alpha,p^\nu_\beta\}=\delta_{\alpha\beta}g^{\mu\nu}.
\end{equation}

\noindent
The Poincar\'e groups acts canonically as:

\begin{equation}
\label{anit15}
x_{\mu\alpha}\longmapsto x'_{\mu\alpha}=\Lambda_\mu^\nu\, x_{\nu\alpha}+a_\mu,\quad p_{\mu\alpha}\longmapsto p'_{\mu\alpha}=\Lambda_{\mu}^\nu\, p_{\nu\alpha},
\end{equation}

\noindent
and its generators are:

\begin{equation}
\label{anit15}
J_{\mu\nu}=\sum_\alpha(x_{\mu\alpha}p_{\nu\alpha}-x_{\nu\alpha}p_{\mu\alpha}),\quad P=\sum_\alpha p_{\mu\alpha}.
\end{equation}

\noindent
Following the previous considerations, we have to choose two independent first class constraints, both invariant under the action of the Poincar\'e group. We consider \cite{22}:

\begin{equation}
\label{anit16}
K_1=p_1^2-m_1^2+V,\quad K_2=p_2^2-m_2^2+V
\end{equation}

\noindent
where $V$ is a common interaction term. Since the constraints have to be invariant under the action of the whole Poincar\'e group, the most general interaction term $V$ must be some function of Lorentz scalars and of the difference $x_1-x_2$ (which are invariants under the action of the Lorentz group and of the Translation group, respectively). Thus, we have that $V=V(\xi)$, where:

\begin{equation}
\label{anit17}\xi=r^2-\frac{(P\cdot r)^2}{p^2},\quad P=p_1+p_2,\quad r=\frac{1}{2}(x_1-x_2).
\end{equation}

\noindent
In analogy with the single particle case, by fixing the value of the constraints $K_1$ and $K_2$ we identify a submanifold $\Sigma\subset T^*\mathbb R^8$. Furthermore, they generate transformations which maps $\Sigma$ into itself. The main difference with respect to the single particle case is that now the action of $K_1$ and $K_2$ foliates $\Sigma$ by bi-dimensional surfaces $S$. This means that, when we project the state of motion on the spacetime, we get bi-dimensional surfaces. Therefore, the pair of world lines which describes the motion of the two particles in the spacetime is not unambiguosly determined. To remove this ambiguity, we have to choose a one-dimensional curve in $S$ and discard the rest of $S$ as being of no physical significance. Such a curve can be specified by choosing another constraint $\chi_1(x,p)$ with no explicit dependence on any parameter and with the only requirement that it is not constant over an $S$. To assign a value of an evolution parameter $\tau$ to each point of $C$ we must set up an explicitly dependence on $\tau$ by fixing another constraint $\chi(x,p,\tau)$. The four constraints:

\begin{equation}
\label{anit18}
\begin{split}
&K_1=p^2_1-m_1^2+V(\xi)\approx0,\quad\chi_1(x,p)\approx0\\
&K_2=p_2^2-m_2^2+V(\xi)\approx0,\quad\chi_2(x,p,\tau)\approx0
\end{split}
\end{equation}

\noindent
define the physical interacting two-particle system. The set of this four constraints is of second class, i.e.:

\begin{equation}
\label{anit19}
det\,|\{\chi_\alpha,K_\beta\}|\neq0.
\end{equation}

\noindent
Thus, we can define Dirac Brackets which support the twelve-dimensional phase space defined by Eqs. (\ref{anit18}) and the state of motion is a curve $C$ on the sheet $S$. Poincar\'e group is realized in a canonical way w.r.t. these DB. Following our discussion in the previous subsection, the WLC is easy to set up. Indeed, the difference w.r.t. the single particle case is that now the condition (\ref{wlc16}) has to be satisfied separataely by each particle.

It is possible to show \cite{22} that if we choose the evolution parameter by setting the constraint $\chi_2(x,p,\tau)$ in a \textit{kinematical} way (e.g., $\chi_2=(x^0_1-x_2^0)$), i.e., if it does not depend on the state of motion of the particles and we require the WLC to be satisfied, the No-Interaction Theorem reappears. Therefore, we have to choose the evolution parameter \textit{dynamically}, i.e. by setting for istance:

\begin{equation}
\label{anit20}
\chi_1=P\cdot r,\quad\chi_2=\frac{1}{2}P\cdot(x_1+x_2)-\tau.
\end{equation}

\noindent
If we now compute the explicit expression of the Dirac Brackets\footnote{See Appendix \ref{appdb} for the explicit computations.} we obtain:

\begin{equation}
\label{28}
\{f,g\}_D=\{f,g\}-\sum_{\alpha\beta}\{f,v_\alpha\}A^{-1}_{\alpha\beta}\{v_\beta,g\}
\end{equation}

\noindent
where the $v$'s runs on the constraints set and the matrix $A^{-1}_{\alpha\beta}$ is defined in (\ref{anit27}). With this choice of the constraints the WLC condition is satisfied \cite{22}. 

By means of this construction, we were able to define a model of two physical relativistic interacting particles on a reduced carrier space on which a canonical realization of the Poincar\'e group is defined. Thus, ipso facto, \textit{we have evaded the No-Interaction Theorem}. It is natural to ask now what hypothesis of the No-Interaction Theorem does not hold in this case. The answer is that in the reduced space the physical positions do not commute anymore w.r.t. the DB (\ref{28}). More precisely, if we consider the canonical projection:

\begin{equation}
\label{anit30}
\xymatrix{
T^*(\mathbb R^4\times\mathbb R^4)\ar[d]^{\pi}\\
\mathbb R^4\times\mathbb R^4
}
\end{equation}

\noindent
we can take the pull-back of the coordinates on the basis of this fiber bundle through the canonical projection and consider their commutation relation w.r.t. the DB, i.e.:

\begin{equation}
\label{anit31}
\{\pi^*(x^\mu_\alpha),\pi^*(x^\nu_\alpha)\}_D\neq0,\quad\alpha=1,2.
\end{equation}

\noindent
Therefore, the price that we have paid to evade the No-Interaction Theorem is that we have lost the localizability of the two particles. It implies that a Lagrangian function for this system cannot exists\footnote{A proof of this proposition is given in Appendix \ref{appf}.} and then the No-Interaction Theorem does not hold\footnote{Remember that the fundamental hypothesis that we have used in Appendix \ref{appb} to prove the No-Interaction Theorem is the existence of Lagrangian (regular or not) for the system.}. 
\\

\noindent
\textbf{Remark: } The consequences of this evasion, e.g. the loss of localizability of the particles, are the fundamental reasons that impose us to pass to a non-approximate description in terms of fields.

\chapter*{Outlook and Conclusions}
\addcontentsline{toc}{chapter}{Outlook and Conclusions}

The main result of this work lies in the fact that, to achieve a covariant description of relativistic particle dynamics in terms of Poisson Brackets, localizability of the particle has been lost. Indeed, as we have shown in Chapter \ref{capitolodue} for both single particle and for many-bodies system, the Poisson description of the presymplectic manifolds which are the carrier spaces for the dynamics of such systems gives rise to a Poisson Brackets in which the positions do not commute anymore. Let us consider for istance the last of Eqs. (\ref{pbd3})) relative to the single relativistic particle case:

\begin{equation}
\label{conc1}
\left\{x_\rho,x_\sigma\right\}=m^2c^2\frac{\dot x_\rho x_\sigma-\dot x_\sigma x_\rho}{\mathscr L}.
\end{equation}

\noindent
Since the r.h.s. of Eq. (\ref{conc1}) is the ratio of a generator of the Lorentz group and of a Casimir function, we can set:

\begin{equation}
\label{conc2}
m^2c^2\frac{\dot x_\rho x_\sigma-\dot x_\sigma x_\rho}{\mathscr L}=\frac{\ell_{\rho\sigma}}{K}
\end{equation}

\noindent
where $K\in\mathbb R$ and $\ell_{\rho\sigma}=-\ell_{\sigma\rho}$. By means of this identification (\ref{conc2}), we can define a Poisson structure on the dual of the Poincar\'e algebra $\mathcal P^*=\mathcal L^*\rtimes {R^4}^*$ as:

\begin{equation}
\label{conc3}
\begin{split}
&\{x_\rho,x_\sigma\}=\frac{\ell_{\rho\sigma}}{K}\\
&\{\ell_{\mu\nu},x_\rho\}=\eta_{\mu\rho}x_\nu-\eta_{\nu\rho}x_\mu\\
&\{\ell_{\mu\nu},\ell_{\rho\sigma}\}=\eta_{\mu\rho}\ell_{\nu\sigma}-\eta_{\mu\sigma}\ell_{\nu\rho}-\eta_{\nu\rho}\ell_{\mu\sigma}+\eta_{\nu\sigma}\ell_{\mu\rho}
\end{split}
\end{equation}

\noindent
It is possible to show that these Poisson Brackets satisfy the Jacobi identity. They reproduces a deformation of the Poincar\'e algebra where positions do not commute anymore but are proportional to the generators of the Lorentz group and in the limit $K\longrightarrow\infty$ one obtains the usual Poincar\'e algebra.

This bracket turn out to be related to what is usually done in non-commutative geometry and to what was done by Snyder \cite{50} to introduce a quantum space-time.

Notice that, if we require that Poisson Brackets are associated with a dimensionless tensor, it means that the non-zero commutator between positions imply the existence of a fundamental length because in this case $\{x^\mu,x^\nu\}$ has the dimension of the square of a length\footnote{It is the analogue of what one usually does in Quantum Mechanics when requires that Poisson Bracket is dimensionless and therefore introduces a fundamental constant with the dimension of an action, i.e. $\hbar$.}. 

Therefore, it is gratifying that our result of non-commuting positions stemming from Poincar\'e covariance suggests the existence of a fundamental length similar to what has been postulated in non-commutative geometry and in the approach of Snyder to quantum space-time.

This result suggests that we are on the right track to describe covariant Poisson Brackets for fields and, eventually, to construct by reduction procedure interacting field.


\appendix
\chapter{Analysis of Space-Time}\label{altraappendice}
In this Appendix we will outline how, by means of the principle of mutual objective existence, the relativity group and the Lorentz metric tensor are derived.

\section{Compatible Frames and Objective Existence}
Let us consider two reference frames $R$ and $R'$ that can be decomposed as $R=\theta\otimes\Gamma$ and $R'=\theta'\otimes\Gamma'$, respectively. We will give now the following:

\begin{defn}
\textbf{(Compatibility condition): }Two reference frames $R$ and $R'$ are called a \textbf{compatible pair} if:

\begin{equation}
\label{rf7}
\text{Tr }(R \cdot R')\neq 0,
\end{equation}
\end{defn}

\noindent
This condition means that two compatible systems will perceive the observers of each other as representing some existing physical object. For this reason, when condition (\ref{rf7}) holds, $R$ and $R'$ are said to satisfy the so called \textit{mutual objective existence} condition:

\begin{equation}
\label{rf8}
\theta(\Gamma') \neq 0, \qquad \qquad \theta'(\Gamma) \neq 0\, \,.
\end{equation}

\noindent
Let us explain the physical meaning of this condition. When Eqs. (\ref{rf8}) hold, the integral curves of the vector field $\Gamma'$ (associated with the reference frame $R'$), which represents the evolution of one observer, is not included into the simultaneity surfaces of the other frame $R$ and viceversa. In other words, by imposing Eqs.(\ref{rf8}), we are excluding the possibility of a time-axis of one reference frame to be contained in the space-axes of the other reference frame. The physical reason is quite obvious. Indeed if these conditions do not hold, then one observer will see the world lines associated with the other observer only at one instant of time (which is that of the leaf we are considering) and so they do not perceive the existence of the other neither in the past nor in the future.
\\

\noindent
\textbf{Remark}: From now on we will consider only the case in which $\theta(\Gamma')>0$ and $\theta'(\Gamma)>0$. Then, the compatibility condition (\ref{rf7}) will be $\text{Tr }(R \cdot R') > 0$. However it should be stressed that the opposite case $(<0)$ still has a physical interpretation in terms of antiparticles which travel backwards in time rather than particles which travel forwards in time. 
\\

We will show now (in two dimensions, for simplicity) that the group connecting pairwise compatible reference frames is necessarily the Lorentz group.

\begin{figure}[h!]
\centering
\includegraphics[scale = 0.55]{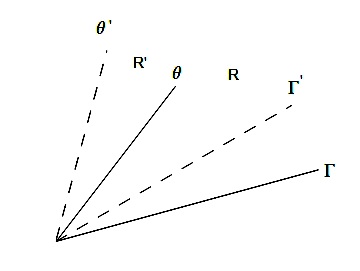}
\caption{\textit{Representation of two comaptible reference frames $R$ and $R'$ in a two-dimensional space-time.}}
\label{bidim}
\end{figure}

\begin{thm}
Let $R$ and $R'$ be two compatible reference frames (i.e. $Tr(R \cdot R')>0$) and let $\varphi$ be a (linear) transformation such that:

\begin{equation}
\theta\overset{\varphi}{\longmapsto}\theta',\quad\Gamma\overset{\varphi}{\longmapsto}\Gamma'
\end{equation}

\noindent
then the transformation $\varphi$ belongs to the Lorentz group.
\end{thm}
\proof
We are considering two compatible reference frames $R$ and $R'$ as shown in Fig \ref{bidim}.

\noindent
The most simple way to prove this theorem is to start with a fiducial reference frame, say $R$, and to consider the subgroup of the inhomogeneous linear group with the requirement, that out of $R$, all transformed reference frames are pairwise compatible. The subgroup selected by implementing the mutual objective existence condition will be the relativity group. 

In a two-dimensional space-time, linear transformations from a reference frame to another will be given by the general linear group $GL(2, \mathbb{R})$. We have to remove dilations because they are only a change of scale and $R'$ would be only a reparametrization of $R$ (i.e., trivial compatibility). Thus, we remain with the special linear group $SL(2, \mathbb{R})$ which is formed by those transformations which preserve the volume. Since in two dimensions the volume form is defined by means of the symplectic structure, $SL(2, \mathbb{R})$ coincides with the symplectic group $Sp(2, \mathbb{R})$. This means that the generators of the transformations will be Hamiltonians and, in order to preserve mutual objective existence, the corresponding Hamiltonian functions will be quadratic. In two dimensions, the quadratic forms can only\footnote{Actually, there is also the case of couple of lines $aX^2,\, b Y^2$ and the case $XY$. We will not treat these cases here because it may keep us too far away from our discussion. We refer to \cite{30, 31, 32} for a detailed treatment.} have the following expression:

\begin{itemize}
\item[i)] $a^2 X^2+b^2 Y^2$
\item[ii)] $a^2 X^2-b^2 Y^2$
\end{itemize}

\noindent
The level sets of these functions are the orbits of the group of transformations that we are looking for. Thus, in two dimensions, they can be an ellipse or an hyperbola. Let us examine these two cases separately.

\begin{itemize}
\item[i)] $a^2 X^2+b^2 Y^2$ is the case of the ellipse. In this case, with reference to Fig. \ref{ellipse}, it will rotate $\theta'$ starting from $\theta$ and it will also overlap with $\Gamma$. But then the mutual objective existence is violated because we rotate until we get the instant of time in which world lines exist only at that instant (i.e. $\theta'(\Gamma) = 0$).

\begin{figure}[h!]
\centering
\includegraphics[scale = 0.5]{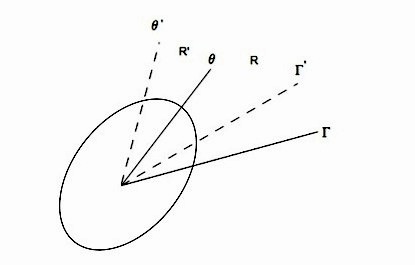}
\caption{\textit{The rotations group does not preserve the mutual objective existence condition}.}
\label{ellipse}
\end{figure}

So we have to exclude positive definite quadratic functions because they would generate rotations and therefore the transformed frame would not be compatible with the starting frame;

\item[ii)] $a^2 X^2-b^2 Y^2$ is the case of the hyperbola. In Fig. \ref{hyp} are shown the orbits of the transformation. It is clear that, if the splitting are oriented as in Fig. \ref{hyp}, then this case we will never violate the mutual objective existence condition because we can move $\Gamma'$ (or $\Gamma$) towards the hyperbola without reaching $\theta$ (or $\theta'$) for any value of the parameter of the transformation. Since the orbits is a hyperbola, we can identify the group that we are looking for with the Lorentz group.

\begin{figure}[h!]
\centering
\includegraphics[scale = 0.65]{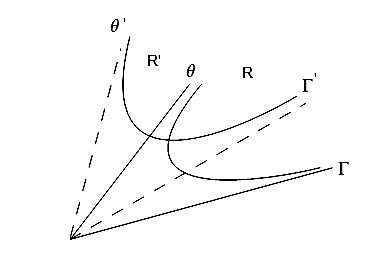}
\caption{\textit{The Lorentz group preserves the mutual objective existence condition}.}
\label{hyp}
\end{figure}

\noindent
Thus, in our approach the Lorentz group emerges as the group which preserves the mutual objective existence, without making any reference to the Lorentz metric tensor. 

The analysis for the four-dimensional space-time, which becomes more cumbersome without teaching us nothing new, can be repeated along the same lines.
\endproof

\end{itemize}

\section{Symmetric Tensors Associated with Equivalence Classes of Reference Frames}
We will show now that the Lorentz metric tensor emerges as a consequence of the covariant formalization of reference frames. Let us consider the family of reference frames in the equivalence class identified by Lorentz transformations. More precisely, by using the Lorentz group, we can generate sets of couples $\bigl(\varphi_g(\alpha),\,\varphi_g(\Gamma)\bigr)$, where $\varphi_g$, with a slight abuse of notation, is the transformation associated with the element $g\in\mathbb G$, where $\mathbb G$ is the Lorentz group. If $\{\varphi_g(\alpha)\}_{g\in\mathbb G}$ and $\{\varphi_g(\Gamma)\}_{g\in\mathbb G}$ are a basis of one-forms and vector fields respectively, we may define a symmetric covariant tensor field $\eta$ by setting:

\begin{equation}
\eta\bigl(\varphi_g(\Gamma)\bigr)=\varphi_g(\alpha)
\end{equation}

\noindent
or, equivalently, a symmetric controvariant tensor:

\begin{equation}
\tilde\eta\bigl(\varphi_g(\alpha)\bigr)=\varphi_g(\Gamma).
\end{equation}

\noindent
It is easy to show (in coordinates) that this tensor is the Lorentz metric tensor:

\begin{equation}
\eta=(dx_0)^2-(d\vec{x})^2=dx_0\otimes dx_0-d\vec x\otimes d\vec x
\end{equation}

\noindent
which is preserved by the action of the Lorentz group. Its contravariant form is

\begin{equation}
\tilde{\eta} = \frac{\partial}{\partial{x_0}}\otimes \frac{\partial}{\partial{x_0}}-\bigg{(}\frac{\partial}{\partial{x}}\otimes \frac{\partial}{\partial{x}}+\frac{\partial}{\partial{y}}\otimes \frac{\partial}{\partial{y}}+\frac{\partial}{\partial{z}}\otimes \frac{\partial}{\partial{z}}\bigg{)}\,,
\end{equation}

\noindent
and they are one the inverse of the other.
\\

\noindent
In conclusion, starting from the covariant definition of reference frames we have shown how the Lorentz group emerges in this setting as the group which preserves the mutual objective existence and how the Lorentz metric tensor is \textit{derived} in this contest.


\chapter{Symplectic and Canonical Formalism on Tangent Bundle}\label{s21}
In this Appendix we will briefly explain how to introduce canonical formalism directly on the Tangent Bundle. It is of interest for us because, once a manifold is given, the Tangent Bundle is naturally specified as the local model space of the manifold, also for infinite dimensions. It is not the case of the Cotangent Bundle, defined as the space of linear functionals over the vectors, because at infinite dimensions it is necessary to consider also the topology of the manifold. However, unlike what happens for the Cotangent Bundle, in which the symplectic structure is canonical, it is not the case of the Tangent Bundle. 

In this Appendix we will show that on the Tangent Bundle it is possible to define a symplectic structure which depends on a Lagrangian function\footnote{In this sense, it is not a canonical structure.}. Moreover, we will study the case of a presymplectic manifold (i.e., a manifold equipped with a degenerate closed 2-form) and we will show how it is possible to define on it Poisson Brackets.
\\

Let us consider an even dimensional and orientable manifold $\mathbb{M}$. We have the following \cite{2}\cite{3}:

\begin{defn}
\textbf{(Symplectic Manifold): }An even-dimensional orientable manifold $\mathbb M$ equipped with a symplectic structure $\omega$ on it is called a \textbf{symplectic manifold} and it is usually denoted by $(\mathbb M,\omega)$.
\end{defn}

\noindent
Let us recall what is a symplectic structure by means of the following \cite{3}:
\begin{defn}
\textbf{(Symplectic Form): }A \textbf{symplectic form} $\omega$ is a non-degenerate and closed two-form $\omega$, i.e. it is:

\begin{itemize}

\item[i)] \textbf{Non-degenerate}:

\begin{equation}
\label{nds143}
\omega(X,Y)=0,\quad\forall Y\Longleftrightarrow X=0;
\end{equation}

\item[ii)] \textbf{Closed} :

\begin{equation}
\label{nds144}
d\omega=0.
\end{equation}

\end{itemize}
\end{defn}

\noindent
Moreover, since $\omega(X,Y)\equiv i_Y i_X\omega$, the non-degeneracy condition (\ref{nds143}) can be written in the following equivalent form:

\begin{equation}
\label{nds145}
i_X\omega=0\Longleftrightarrow X=0.
\end{equation}

\noindent
\textbf{Remark: }A symplectic form on a manifold establishes a bijection between vector fields and $1$-form. Indeed, by taking the contraction of $\omega$ with a vector field $X$ on $\mathbb{M}$ we define a $1$-form $\alpha=i_X\omega\in\mathfrak X^*(\mathbb{M})$. Viceversa, given a $1$-form $\alpha$ on $\mathbb{M}$, in virtue of the non-degeneracy of $\omega$, the equation $\alpha=i_X\omega$ uniquely determines the vector field $X\in\mathfrak X(\mathbb{M})$.

Let now consider a function $\mathscr{L}$ on $\in\mathcal F (TQ)$\footnote{For the time being, this function $\mathscr{L}\in\mathcal F(TQ)$ is not necessarily the Lagrangian function associated with a second-order vector field $\Gamma$.} and the associated Cartan form $\theta_\mathscr{L}=\frac{\partial\mathscr{L}}{\partial{v^i}}{dq^i}$. We define the following two-form on $TQ$:

\begin{equation}
\label{nds146}
\omega_\mathscr{L}:=-d\theta_\mathscr{L}.
\end{equation}

\noindent
By using the local coordinate expression $\theta_\mathscr{L}$, we easily find the expression of the $2$-form (\ref{nds146}) in a local coordinate system as follows:

\begin{equation}
\label{nds147}
\begin{split}
\omega_\mathscr{L}&=-d\left(\frac{\partial{\mathscr{L}}}{\partial{v^i}}dq^i\right)\\
&=-d\left(\frac{\partial{\mathscr{L}}}{\partial{v^i}}\right)\wedge dq^i\\
&=-\left(\frac{\partial^2{\mathscr{L}}}{\partial{v^i}\partial{v^j}}dv^j\wedge dq^i+\frac{\partial^2{\mathscr{L}}}{\partial{v^i}\partial{q^j}}dq^j\wedge dq^i\right)\\
&=\frac{\partial^2{\mathscr{L}}}{\partial{v^i}\partial{v^j}}dq^i\wedge dv^j+\frac{1}{2}\left(\frac{\partial^2{\mathscr{L}}}{\partial{v^i}\partial{q^j}}-\frac{\partial^2{\mathscr{L}}}{\partial{v^j}\partial{q^i}}\right)dq^i\wedge dq^j.
\end{split}
\end{equation}

\noindent
The following Propositions list the properties satisfied by the $2$-form $\omega_\mathscr{L}$ defined in Eq. (\ref{nds146}).

\begin{prop}\label{p21}
The $2$-form $\omega_\mathscr{L}=-d\theta_\mathscr{L}$ whose local coordinate expression is given by (\ref{nds147}) has the following properties:

\begin{itemize}

\item[i)]$\omega_\mathscr{L}$ is closed (i.e., $d\omega_\mathscr{L}=0$);

\item[ii)]$\omega_\mathscr{L}$ does not contain terms proportional to $dv^i\wedge dv^j$;

\item[iii)] $L_\Gamma\omega_\mathscr{L}=0$ for a second-order vector field $\Gamma$.

\end{itemize}
\end{prop}

\begin{proof}
The first property follows directly from the definition (\ref{nds146}) which shows that $\omega_{\mathscr L}$ is an exact $2$-form. As regard the second property, by construction $\omega_\mathscr{L}$ does not contain terms $dv^i\wedge dv^j$ as is evident from its local coordinate expression (\ref{nds147}). Finally, by taking $\Gamma$ as a solution of the Euler - Lagrange equations:

\begin{equation}
\label{nds148}
L_\Gamma\theta_\mathscr{L}-d\mathscr{L}=0
\end{equation}

\noindent
and taking the exterior derivative of these equations, we have:

\begin{equation}
\label{nds149}
d(L_\Gamma\theta_\mathscr{L}-d\mathscr{L})=0\quad\rightarrow\quad dL_\Gamma \theta_\mathscr{L}=0
\end{equation}

\noindent
from which, by using Cartan's identity $L_\Gamma=i_\Gamma d+d i_\Gamma$ and the property $d^2=0$ we obtain:

\begin{equation}
\label{nds150}
0=d i_\Gamma d\theta_\mathscr{L}=-di_\Gamma\omega_\mathscr{L}=L_\Gamma\omega_\mathscr{L}
\end{equation}

\noindent
where in the last equality we have used property i). This proves the property iii) and closes the proof of the proposition.
\end{proof}

\begin{prop}
The $2$-form $\omega_\mathscr{L}$ associated to $\mathscr{L}\in\mathcal F(TQ)$ and defined in Eq. (\ref{nds146}) is symplectic iff $\mathscr{L}$ is regular.
\end{prop}

\begin{proof}
Let $X$ be a vector field on $TQ$, i.e.:

\begin{equation}
\label{nds151}
X=f^i\frac{\partial}{\partial{q^i}}+g^i\frac{\partial}{\partial{v^i}}\in\mathfrak X(TQ).
\end{equation}

\noindent
Then, by using the coordinates expression (\ref{nds147}) of $\omega_\mathscr{L}$, we find by direct computation:

\begin{equation}
\label{nds152}
i_X\omega_\mathscr{L}=\frac{\partial^2{\mathscr{L}}}{\partial{v^i}\partial{v^j}}f^jdv^i+\left[\frac{\partial^2{\mathscr{L}}}{\partial{v^i}\partial{v^j}}g^j+\left(\frac{\partial^2{\mathscr{L}}}{\partial{v^i}\partial{q^j}}-\frac{\partial^2{\mathscr{L}}}{\partial{v^j}\partial{q^i}}\right)f^j\right]dq^i.
\end{equation}

\noindent
Therefore, the non-degeneracy condition (\ref{nds145}) for $\omega_\mathscr{L}$ is equivalent to the condition of regularity of $\mathscr L$:

\begin{equation}
\label{nds153}
\text{det }\biggl|\biggl|\frac{\partial^2{\mathscr{L}}}{\partial{v^i}\partial{v^j}}\biggr|\biggr|\neq0.
\end{equation}

\noindent
Indeed, the non-degeneracy condition means that $i_X\omega_\mathscr{L}=0$ iff $X=0$. Thus, if $X=0$ it has vanishing components ($f^i=g^i=0$ in Eq. (\ref{nds151})) and the determinant of the Hessian is different from zero and $\mathscr L$ is regular. Viceversa, if $\omega_{\mathscr{L}}$ is degenerate the determinant of the Hessian must be zero. Therefore, due to the closure of $\omega_\mathscr{L}$, saying that it is a closed non-degenerate $2$-form (i.e., a symplectic form) is equivalent to say that the function $\mathscr{L}\in\mathcal F(TQ)$ associated with $\omega_\mathscr{L}$ is regular.
\end{proof}

What we have said up to now is true for a generic function $\mathscr{L}$ on $TQ$ without making reference to any dynamical problem. Let us now consider the case in which $\mathscr{L}$ is exactly a regular Lagrangian function on $TQ$. The previous construction allows us to define a \textit{Lagrangian symplectic structure} and therefore the Tangent Bundle $TQ$ becomes a symplectic manifold $(TQ,\omega_{\mathscr L})$. Since we have a symplectic structure, it is natural to ask if it is possible to define the Hamiltonian formalism directily on the Tangent Bundle. Let us start with the Euler-Lagrange equations in their intrinsic form:

\begin{equation}
\label{nds154}
L_\Gamma\theta_\mathscr{L}-d\mathscr{L}=0.
\end{equation}

\noindent
By using Cartan's identity\footnote{We stress that it is essentially the integration by part at an infinitesimal level.}

\begin{equation}
\label{nds155}
L_\Gamma=i_\Gamma d+di_\Gamma
\end{equation}

\noindent
we have

\begin{equation}
\label{nds156}
i_\Gamma d\theta_\mathscr{L}+di_\Gamma\theta_\mathscr{L}-d\mathscr{L}=0
\end{equation}

\noindent
i.e.:

\begin{equation}
\label{nds157}
i_\Gamma d\theta_\mathscr{L}=d(\mathscr{L}-i_\Gamma\theta_\mathscr{L}).
\end{equation}

\noindent
Now, if we express the quantity inside the brackets in local coordinates, by using the expressions:

\begin{equation}
\label{nds158}
\Gamma=v^i\frac{\partial}{\partial{q^i}}+F^i\frac{\partial}{\partial{v^i}}
\end{equation}

\begin{equation}
\label{nds159}
\theta_\mathscr{L}=\frac{\partial{\mathscr{L}}}{\partial{v^j}}dq^j
\end{equation}

\noindent
we find that:

\begin{equation}
\label{nds160}
\mathscr{L}-i_\Gamma\theta_\mathscr{L}=\mathscr{L}-v^i\frac{\partial{\mathscr{L}}}{\partial{v^i}}.
\end{equation}

\noindent
But this is exactly the opposite of the lagrangian energy $E_\mathscr{L}$ associated with the Lagrangian $\mathscr{L}$ \cite{3}. Thus, by defining

\begin{equation}
\label{nds161}
E_\mathscr{L}\equiv i_\Gamma\theta_\mathscr{L}-\mathscr{L}
\end{equation}

\noindent
we see from the definition (\ref{nds146}) of $\omega_\mathscr{L}$ that Eqs. (\ref{nds157}) can be written in the form:

\begin{equation}
\label{nds162}
i_\Gamma\omega_\mathscr{L}=-dE_\mathscr{L}.
\end{equation}

\noindent
This is another form of the Euler-Lagrange equation which can be regarded as an algebraic equation for the second-order vector field $\Gamma$. The solution exists and is unique if and only if $\omega_\mathscr{L}$ is non-degenerate (i.e., iff $\mathscr{L}$ is regular). Moreover, Eq. (\ref{nds162}) is of the Hamiltonian form. Indeed, let us recall that an \textbf{Hamiltonian vector field} $X_H$ associated to a given smooth function $H$ on a symplectic manifold $(\mathbb M,\omega)$ is the unique vector field $X_H$ such that

\begin{equation}
i_{X_H}\omega=dH.
\end{equation}

\noindent
From Cartan's identity ($L_{X_H}=i_{X_H}d+di_{X_H}$) and the closure property of $\omega$ ($d\omega=0$) follows taht $L_{X_H}\omega=0$ for any Hamiltonian vector field. Vice-versa, given a vector field $X\in\mathfrak X(\mathbb M)$ such that $L_X\omega=0$, then $di_X\omega=0$ and so by Poincare's Lemma there exists (at least locally) a function $H\in\mathcal F(\mathbb M)$ such that $i_X\omega=dH$, i.e., $X=X_H$. A vector field $X$ on $\mathbb M$ satisfying the condition $L_X\omega=0$ is called a \textit{locally Hamiltonian vector field}. Hence, in the case $(TQ,\omega_\mathscr{L})$ we see from Eq. (\ref{nds162}) that $\Gamma$ is the Hamiltonian vector field associated to $E_\mathscr{L}$. Therefore we say that Eq. (\ref{nds162}) together with Eq. (\ref{nds161}) define the \textit{Hamiltonian (or Canonical) formalism directly on the tangent bundle $TQ$} without any need to go to the phase space $T^*Q$.

\noindent
As an example, let us consider the following Lagrangian function:

\begin{equation}
\label{nds164}
\mathscr{L}=\frac{1}{2}\delta_{ij}v^iv^j-U(q).
\end{equation}

\noindent
Then, from Eq. (\ref{nds149}) and Eq. (\ref{nds161}), we have respectively:

\begin{equation}
\label{nds165}
\omega_\mathscr{L}=\delta_{ij}dq^i\wedge dv^j
\end{equation}

\begin{equation}
\label{nds166}
E_\mathscr{L}=\delta_{ij}v^jdv^i+dU(q)=\delta_{ij}v^jdv^i+\frac{\partial{U}}{\partial{q^j}}dq^j.
\end{equation}

\noindent
Therefore, Eq. (\ref{nds162}) implies that:

\begin{equation}
\label{nds169}
\Phi^i=-\delta^{ij}\frac{\partial{U}}{\partial{q^j}}=-\frac{\partial{E_\mathscr{L}}}{\partial{q^i}}
\end{equation}

\noindent
The equations of motion become:

\begin{equation}
\label{nds170}
\begin{cases}
\frac{dq^i}{dt}=v^i=\frac{\partial{E_\mathscr{L}}}{\partial{v^i}}\\
\frac{dv^i}{dt}=\Phi^i=-\frac{\partial{E_\mathscr{L}}}{\partial{q^i}}
\end{cases}\Rightarrow\quad
\begin{cases}
\frac{dq^i}{dt}=\frac{\partial{E_\mathscr{L}}}{\partial{v^i}}\\
\frac{dv^i}{dt}=-\frac{\partial{E_\mathscr{L}}}{\partial{q^i}}
\end{cases}
\end{equation}

\noindent
which present the same structure of Hamilton's canonical equations but written in the variables $(q^i,v^i)$ on $TQ$ and not in the usual positions and (conjugate) momenta.

\section{Poisson Brackets on Symplectic Manifolds}\label{s211}
What we would like to show in this section is that it is possible to construct Poisson Brackets (PB) directly on $TQ$ in the case of non-degenerate Lagrangian 2-form. We will proceed in the following way: we first consider the case of a generic symplectic manifold and then we specialize our treatment to the case of the Tangent Bundle $TQ$.

Let us consider a generic symplectic manifold $(\mathbb M,\omega)$. We can define PB on it in the following way:

\begin{itemize}

\item Given $f,g\in\mathcal F(\mathbb M)$, we solve $i_{X_f}\omega=df$ and $i_{X_g}\omega=dg$ to get the vector fields $X_f$ and $X_g$ associated with $f$ and $g$, respectively. In other words, given two (smooth) functions $f$ and $g$ on $\mathbb M$, we consider the Hamiltonian vector field $X_f$ , $X_g$ associated with them\footnote{Let us recall that the definition of Hamiltonian vector fields requires the non-degeneracy of the $2$-form $\omega$.};

\item The map:

\begin{equation}
\label{nds179}
\{\cdot,\cdot\}\,:\,\mathcal F(\mathbb M)\times\mathcal F(\mathbb M)\longrightarrow \mathcal F(\mathbb M)
\end{equation}

\noindent
with

\begin{equation}
\label{nds180}
(f,g)\longmapsto\{f,g\}:=\omega(X_f,X_g)
\end{equation}

\noindent
defines a Poisson Bracket\footnote{Poisson Brackets defined in this way are often called Lagrangian Brackets. The "inverse" of this expression, i.e. $\omega(\Lambda)=\mathds1$ defines the so-called Poisson Tensor.}.

\end{itemize}

\noindent
To proof that the map (\ref{nds180}) effectively defines a PB we have to verify that it satisfies all the properties of a PB, i.e.:

\begin{defn}
Let $(\mathbb M,\omega)$ be a symplectic manifold. The PB on $\mathbb M$ defined by (\ref{nds180}) enjoys the following properties:

\begin{itemize}

\item[i)] \textbf{Skew-symmetry:}

\begin{equation}
\label{nds181a}
\{f,g\}+\{g,f\}=0,\quad\forall f,g\in\mathcal F(\mathbb M);
\end{equation}

\item[ii)] \textbf{Bilinearity:}

\begin{equation}
\label{nds181b}
\{\alpha f+\beta g,h\}=\alpha\{f,h\}+\beta\{g,h\},\quad\forall f,g,h\in\mathcal F(\mathbb M),\,\alpha,\beta\in\mathbb R;
\end{equation}

\item[iii)]\textbf{Jacobi identity:}

\begin{equation}
\label{nds181d}
\{f,\{g,h\}\}+\{g,\{h,f\}\}+\{h,\{f,g\}\}=0,\quad\forall f,g,h\in \mathcal F(\mathbb M).
\end{equation}

\item[iv)] \textbf{Leibniz rule\footnote{This condition defines the compatibility between the Lie product and the associative and commutative product between functions. If this product is not commutative, we obtain the so-called q-Poisson Bracket.}:}

\begin{equation}
\label{nds181c}
\{f,gh\}=\{f,g\}h+g\{f,h\},\quad\forall f,g,h\in\mathcal F(\mathbb M);
\end{equation}

\end{itemize}
\end{defn}

\begin{proof}
The property i) follows directly by the definition (\ref{nds180}) of $\{\cdot,\cdot\}$ and by the skew-symmetry of $\omega$. As regard property ii), it also derives from the definition and the fact that the association $f\longmapsto X_f$ is linear in $f$. To prove property iii) we first observe that the map (\ref{nds180}) can be written in the following equivalent forms:

\begin{equation}
\label{nds182}
\{f,g\}=\omega(X_f,X_g)=i_{X_g}i_{X_f}\omega=i_{X_g}df=-i_{X_f}dg=L_{X_g}f=-L_{X_f}g.
\end{equation}

\noindent
Therefore:

\begin{equation}
\label{nds183}
\{f,gh\}=-L_{X_f}(gh)
\end{equation}

\noindent
and property iii) is immediately proved by using the Leibniz rule in the r.h.s. of (\ref{nds183}) and again the last equality of Eqs. (\ref{nds182}). As for iv) Pauli proved it by showing the equivalence between the Jacobi identity for the PB and the closure of the symplectic structure \cite{49}.
\end{proof}

\noindent
\textbf{Remark: } Up to now we have not defined any dynamics on $\mathbb M$ and we have discussed PB on symplectic manifolds from the geometrical point of view. Let us now consider a function $H\in\mathcal F(\mathbb M)$ as the Hamiltonian and let $X_H$ be the Hamiltonian vector field associated with it, i.e., $i_{X_H}\omega=dH$. Now, if $X_f$ is the Hamiltonian vector field associated with a given $f\in\mathcal F(\mathbb M)$, we have:

\begin{equation}
\label{nds192}
L_{X_H}f=i_{X_h}df=i_{X_H}i_{X_f}\omega=\omega(X_f,X_H),
\end{equation}

\noindent
i.e.:

\begin{equation}
\label{nds193}
L_{X_H}f=\{f,H\}.
\end{equation}

\noindent
Since along the integral curves of $X_H$ we have $L_{X_H}=\frac{d}{dt}$, we find

\begin{equation}
\label{nds194}
\frac{df}{dt}=\{f,H\}
\end{equation}

\noindent
i.e., $f$ obeys the canonical equation of motion in PB notation or equivalently, $X_H$ is the dynamic vector field. Moreover, form Eqs. (\ref{nds193}) and (\ref{nds194}) it follows the well-known result \cite{1}\cite{2} that constants of motion have vanishing PBs with the Hamiltonian and vice-versa. This, together with the Jacobi identity, implies that the PB of any two constants of the motion is itself a constant of the motion.
\\

Let us now focus on the case of the Tangent Bundle $TQ$. If $\omega_\mathscr{L}=-d\theta_\mathscr{L}$ is not degenerate (i.e., if $\mathscr{L}$ is regular), the map

\begin{equation}
\label{nds195}
\{\cdot,\cdot\}_\mathscr{L}\,:\,\mathcal F(TQ)\times\mathcal F(TQ)\longrightarrow\mathcal F(TQ)
\end{equation}

\begin{equation}
\label{nds196}
\{f,g\}_\mathscr{L}=\omega_\mathscr{L}(X_j,X_g),\quad(\text{with }i_{X_f}\omega_\mathscr{L}=df\,\text{and }i_{X_f}\omega_\mathscr{L}=dg)
\end{equation}

\noindent
defines the PB directly on $TQ$.
\\

\noindent
\textbf{Remark: }We would like to stress that since the symplectic $2$-form $\omega_\mathscr{L}$ depends on $\mathscr{L}$, the PB defined in Eq. (\ref{nds196}) also depends on the Lagrangian function $\mathscr{L}$ (this is the reason why we have used the subscript $\mathscr{L}$ in the notation $\{\cdot,\cdot\}_\mathscr{L}$). Therefore, if a given dynamical system admits alternative Lagrangians, then different PBs are derived. Clearly, this has important consequences on quantization if we use the Dirac's prescription of replacing PBs with commutators \cite{14}.
\\

Let us now find the local coordinate expression of the PB defined in Eq. (\ref{nds196}). The vector fields $X_f$  and $X_g$ on $TQ$ are generally represented as:

\begin{equation}
\label{nds197}
X_f=X_f^i\frac{\partial}{\partial{q^i}}+Y^i_f\frac{\partial}{\partial{v^i}}
\end{equation}

\begin{equation}
\label{nds198}
X_g=X^i_g\frac{\partial}{\partial{q^i}}+Y^i_g\frac{\partial}{\partial{v^i}}.
\end{equation}

\noindent
Then, by using the expression (\ref{nds152}), we find:

\begin{equation}
\label{nds199}
\begin{split}
i_{X_f}\omega_\mathscr{L}&=df\Rightarrow\frac{\partial^2{\mathscr{L}}}{\partial{v^i}\partial{v^j}}X^j_f dv^i-\left[\frac{\partial^2{\mathscr{L}}}{\partial{v^i}\partial{v^j}}Y^j_f+\left(\frac{\partial^2{\mathscr{L}}}{\partial{v^i}\partial{q^j}}-\frac{\partial^2{\mathscr{L}}}{\partial{v^j}\partial{q^i}}\right)X^j_f\right]dq^i\\
&=\frac{\partial{f}}{\partial{q^k}}dq^k+\frac{\partial{f}}{\partial{v^k}}dv^k
\end{split}
\end{equation}

\noindent
i.e.:

\begin{equation}
\label{nds200}
\begin{cases}
\frac{\partial^2{\mathscr{L}}}{\partial{v^i}\partial{v^j}}X^j_f=\frac{\partial{f}}{\partial{v^i}}\\
\frac{\partial^2{\mathscr{L}}}{\partial{v^i}\partial{v^j}}Y^j_f+\left(\frac{\partial^2{\mathscr{L}}}{\partial{v^i}\partial{q^j}}-\frac{\partial^2{\mathscr{L}}}{\partial{v^j}\partial{q^i}}\right)X^j_f=-\frac{\partial{f}}{\partial{q^i}}
\end{cases}
\end{equation}

\noindent
and the same for $g$. Therefore, we find:

\begin{equation}
\label{nds201}
\begin{split}
\{f,g\}_\mathscr{L}&:=\omega_\mathscr{L}(X_f,X_g)=i_{X_g}i_{X_f}\omega_\mathscr{L}=X_g^i\frac{\partial{f}}{\partial{q^i}}+Y^i_g\frac{\partial{f}}{\partial{v^i}}\\
&=-X^i_g\frac{\partial{\mathscr{L}}}{\partial{v^i}\partial{v^j}}Y^j_f-X^i_g\left(\frac{\partial^2{\mathscr{L}}}{\partial{v^i}\partial{q^j}}-\frac{\partial^2{\mathscr{L}}}{\partial{v^j}\partial{q^i}}\right)X^j_f+Y^i_g\frac{\partial^2{\mathscr{L}}}{\partial{v^i}\partial{v^j}}X^j_f.
\end{split}
\end{equation}

\noindent
Now, we consider the following three cases:

\begin{itemize}

\item[1)] $f\equiv\frac{\partial{\mathscr{L}}}{\partial{v^j}}$, $g\equiv q^k$. Then:

\begin{equation}
\label{nds202}
\begin{cases}
\frac{\partial{f}}{\partial{v^i}}=\frac{\partial^2{\mathscr{L}}}{\partial{v^i}\partial{v^j}}\,\overset{(\ref{nds200}.a)}{\longrightarrow}\\
\frac{\partial{f}}{\partial{q^i}}=\frac{\partial^2{\mathscr{L}}}{\partial{v^j}\partial{q^i}}\,\overset{(\ref{nds200}.b)}{\longrightarrow}
\end{cases}
\begin{cases}
X^j_f=1,\quad\forall j\\
\frac{\partial^2{\mathscr{L}}}{\partial{v^i}\partial{v^j}}Y^j_f=-\frac{\partial^2{\mathscr{L}}}{\partial{v^i}\partial{q^j}}
\end{cases}
\end{equation}

\noindent
and

\begin{equation}
\label{nds203}
\begin{cases}
\frac{\partial{g}}{\partial{v^i}}=0\,\overset{(\ref{nds200}.a)}{\longrightarrow}\\
\frac{\partial{g}}{\partial{q^i}}=\delta^k_i\,\overset{(\ref{nds200}.b)}{\longrightarrow}
\end{cases}
\begin{cases}
X^j_g=0,\quad\forall j\\
\frac{\partial^2{\mathscr{L}}}{\partial{v^i}\partial{v^j}}Y^j_g=-\delta^k_i
\end{cases}
\end{equation}

\noindent
where we have used the nondegeneracy of $\mathscr{L}$. By substituting the Eqs. (\ref{nds202}) and (\ref{nds203}) into Eq. (\ref{nds201}), we obtain:

\begin{equation}
\label{nds204}
\left\{\frac{\partial{\mathscr{L}}}{\partial{v^j}},q^k\right\}_\mathscr{L}=\delta^k_j;
\end{equation}

\item[2)] $f\equiv q^j$, $g\equiv q^k$. In this case we find the result (\ref{nds203}) for the components of both $X_f$ and $X_g$. Thus, from Eq. (\ref{nds201}) we now find:

\begin{equation}
\label{nds205}
\{q^j,q^k\}_\mathscr{L}=0;
\end{equation}

\item[3)] $f\equiv\frac{\partial{\mathscr{L}}}{\partial{v^j}}$, $g\equiv\frac{\partial{\mathscr{L}}}{\partial{v^k}}$. In this case we find the result (\ref{nds202}) for the components of both $X_f$ and $X_g$. Thus, from Eq. (\ref{nds201}) we find

\begin{equation}
\label{nds206}
\left\{\frac{\partial{\mathscr{L}}}{\partial{v^j}},\frac{\partial{\mathscr{L}}}{\partial{v^k}}\right\}_\mathscr{L}=0.
\end{equation}

\end{itemize}

\noindent
Hence, if we define the momenta as $p_j=\frac{\partial{\mathscr{L}}}{\partial{v^j}}$ (i.e., $p_j$ is only a different name for $\frac{\partial{\mathscr{L}}}{\partial{v^j}}$ and so it is still a function of $q$ and $v$), the Eqs. (\ref{nds204}),(\ref{nds205}) and (\ref{nds206}) give the well-known \textit{canonical commutation relations}:

\begin{equation}
\label{nds207}
\begin{split}
&\{p_j,q^k\}_\mathscr{L}=\delta^k_j\\
&\{p_j,p_k\}_\mathscr{L}=0\\
&\{q^j,q^k\}_\mathscr{L}=0.
\end{split}
\end{equation}

\noindent
These are the \textit{PB that we usually write in terms of position and momenta} on the phase space but, being $p$ a function of $q$ and $v$, they are \textit{written directly on $TQ$} without passing to the Cotangent Bundle which is problematic in infinite dimensions. 

\section{Poisson Brackets on Presymplectic Manifolds}\label{s212}
In the previous section we have showed how to define canonical formalism on Tangent Bundle starting from a regular Lagrangian function. However, despite it is the general case, it is not case in which physicists are interested. Indeed, several physically interesting theories, such as theories of fundamental interactions, are described in terms of non-regular Lagrangians because it is the only way to accomodate constraints and deal with gauge theories. Therefore in this subsection we will show how it is possible to define PB on Presymplectic Manifolds, i.e., manifolds equipped with a closed and degenerate 2-form $\omega$.
\\

Let us go back to the canonical PBs on $TQ$ (\ref{nds204}), (\ref{nds205}) and (\ref{nds206}). As the PB  $\{\cdot,\cdot\}_\mathscr{L}$ acts as a derivation w.r.t. the pointwise associative product (i.e., it satisfies Leibniz rule (\ref{nds181c})), if we expand the derivation $\frac{\partial{\mathscr{L}}}{\partial{v^j}}$ we find that (\ref{nds204}) and (\ref{nds206}) can be written respectively in the form:

\begin{equation}
\label{nds213}
\frac{\partial^2{\mathscr{L}}}{\partial{v^j}\partial{v^m}}\{v^m,q^k\}_\mathscr{L}=\delta^k_j
\end{equation}

\begin{equation}
\label{nds214}
\frac{\partial^2{\mathscr{L}}}{\partial{v^j}\partial{q^m}}\left\{q^m,\frac{\partial{\mathscr{L}}}{\partial{v^k}}\right\}_\mathscr{L}+\frac{\partial^2{\mathscr{L}}}{\partial{v^j}\partial{v^m}}\left\{v^m,\frac{\partial{\mathscr{L}}}{\partial{v^k}}\right\}_\mathscr{L}=0.
\end{equation}

\noindent
By looking Eq. (\ref{nds213}), we easily see that since the product of the two matrices on the l.h.s. gives the identity matrix, then the Lagrangian cannot be degenerate. Indeed, by taking the determinant of both sides of Eq. (\ref{nds213}) and using the Binet Theorem on the l.h.s., we get:

\begin{equation}
\label{nds215}
\text{det }\biggl|\biggl|\frac{\partial^2{\mathscr{L}}}{\partial{v^j}\partial{v^m}}\biggr|\biggr| \cdot\text{det }\Bigl|\Bigl|\{v^m,q^k\}_\mathscr{L}\Bigr|\Bigr|=1.
\end{equation}

\noindent
From which, due to the nondegeneracy of $\omega_\mathscr{L}$ and of $\{\cdot,\cdot\}_\mathscr{L}$, we find

\begin{equation}
\label{nds216}
\text{det }\biggl|\biggl|\frac{\partial^2{\mathscr{L}}}{\partial{v^j}\partial{v^m}}\biggr|\biggr|\neq0.
\end{equation}

\noindent
Therefore, if $\mathscr{L}$ is regular, we can multiply Eq. (\ref{nds213}) by the inverse of the Hessian and we get the PB for the variables parametrizing $TQ$, i.e., between $q$ and $v$. However, if $\mathscr{L}$ is degenerate, we cannot invert Eq. (\ref{nds213}) and we cannot solve it for the PB to get the canonical commutation relation\footnote{At the level of the Euler-Lagrange equations it implies that they are \textit{implicit} differential equations. If we want to formulate the evolution of a quantum system in terms of a one-parameter group of transformations, we need to pass to Hamilton equations, which are \textit{explicit}, and therefore are expressed in terms of a vector field which may be the generator of such transformation. This is the origin of the well-known Dirac-Bergmann Theory of Constraints.}.

Let us show how to define PB on a presymplectic manifold $(M,\omega)$. In this case, $\omega$ is a closed, degenerate 2-form, i.e. $Ker\,\omega\neq\emptyset$. Assume that this kernel $\mathfrak X_\omega=\{X\in\mathfrak X(M)\,|\,i_X\omega=0\}$ defines a regular foliating distribution, i.e. $M$ is foliated in such a way that the natural projection of the fiber bundle $M\overset{\pi}{\longrightarrow}\Sigma$ is a smooth map, where $\Sigma=M/Ker\,\omega$ is the quotient manifold. By using this projection we can define a symplectic 2 form $\omega_\Sigma$ on $\Sigma$ as:

\begin{equation}
\label{pre1}
\pi^*\omega_\Sigma=\omega.
\end{equation}

\noindent
Since $\omega_\Sigma$ is non degenerate, it is possible to define on $\Sigma$ a Poisson structure $\{,\}_\Sigma$ associated with a bivector field $\Lambda_\Sigma=a^{ij}\frac{\partial}{\partial\xi^i}\wedge\frac{\partial}{\partial\xi^j}$. The problem of defining a Poisson structure on the presymplectic manifold $M$ can be now stated in saying that we look for all Poisson structures on $M$, say $\{,\}_M$, such that:

\begin{equation}
\label{pre2}
\{\pi^*f,\pi^*g\}_M=\pi^*\{f,g\}_\Sigma,\quad\forall f,g\in\mathcal F(\Sigma).
\end{equation}

\noindent
It is not difficult to show \cite{15} that bivector fields $\Lambda_M$ giving rise to such Poisson structures can be obtained by taking the linear map $A:\mathfrak X(\Sigma)\longrightarrow\mathfrak X(M)$ (it is rather a 1-1 tensor on $M$) and setting\footnote{We recall that a vector field $X_M$ on $M$ is called $\pi$-projectable if a vector field $X_\Sigma$ on $\Sigma$ exists such that $L_{X_M}\pi^*f=\pi^*L_{X_\Sigma}f$, $\forall f\in\mathcal F(\Sigma)$.}:

\begin{equation}
\label{pre3}
\Lambda_M=a^{ij}A\left(\frac{\partial}{\partial\xi^i}\right)\wedge A\left(\frac{\partial}{\partial\xi^j}\right).
\end{equation}

\noindent
The map $A$ has the following properties:

\begin{equation}
\label{pre4}
A^2=A,\qquad ker\,A=ker\,\omega
\end{equation}

\noindent
and then it is a generalized connection for the fiber bundle $M\overset{\pi}{\longrightarrow}\Sigma$. Now, we can give the following:

\begin{defn}
\textbf{(Compatible Poisson Structure): }A Poisson structure $\Lambda_M$ on $M$ is called compatible with a presymplectic structure $\omega$ on it if:

\begin{equation}
\label{pre5}
\Lambda_M\omega\Lambda_M=\Lambda_M
\end{equation}

\noindent
and

\begin{equation}
\label{pre6}
ker\,\Lambda_M\cap Im\,\omega=\emptyset.
\end{equation}
\end{defn}

\noindent
We will call any such $\Lambda_M$ compatible with $\omega$ a \textit{Poisson Bracket on a presymplectic manifold} $(M,\omega)$.

\noindent
It is possible to prove \cite{15} the following:
\\

\noindent
\textbf{Proposition: }The Poisson Bracket $\{,\}_M$ determined by a presymplectic structure $\omega$ and by a generalized connection $A$, satisfies the Jacobi identity iff the connection is flat.
\\

\noindent
\textbf{Remark: }The existence of a globally defined Poisson Bracket puts some topological conditions on the space. Indeed, if the fiber bundle $M\overset{\pi}{\longrightarrow}M/Ker\,\omega$ is a nontrivial fiber bundle, then there are no flat connections on $M$ and therefore it is impossible to define a global Poisson Bracket on $M$ compatible with the projection $\pi$ \footnote{Consider for example the Hopf fibration $S^3\longrightarrow S^2$.}.
\\

Let us now return at the case of the Tangent Bundle, i.e. we consider the presymplectic manifold $(TQ,\omega_\mathscr L)$. In this case the Lagrangian $\mathscr L=\mathscr L(q,\dot q)$ is not regular, i.e.:

\begin{equation}
\label{pre7}
rank\,\left(\frac{\partial^2\mathscr L}{\partial{\dot q}^i\partial{\dot q}^j}\right)=2h,
\end{equation}

\noindent
where $2h<n=dim\,Q$, $Q$ being the configuration space. It means that the Euler-Lagrange equations:

\begin{equation}
\label{pre8}
i_\Gamma\omega_\mathscr L=-dE_\mathscr L
\end{equation}

\noindent
do not determine a dynamical system on $TQ$, i.e. it is not possible to solve Eqs. (\ref{pre8}) for a vector field $\Gamma$ such that:

\begin{equation}
\label{pre9}
\Gamma=\frac{dq^i}{dt}\frac{\partial}{\partial q^i}+\frac{d{\dot q}^j}{dt}\frac{\partial}{\partial {\dot q}^j}.
\end{equation}

\noindent
However, if we follow the previous construction, it is possible to define a PB on TQ associated with the presymplectic structure $\omega_\mathscr L$ and then we can find a vector field $\Gamma$ associated with the Lagrangian energy $E_\mathscr L$. This dynamical system will be defined on a \textit{reduced surface} of $TQ$ defined by the constraints associated with the degenerate Lagrangian $2$-form $\omega$. If such a vector field $\Gamma$ exists, we say that it provides an \textit{Hamiltonian regularization} of the Euler-Lagrange equation. Therefore, we will conclude this Appendix with the following:
\\

\noindent
\textbf{Proposition: }The Euler-Lagrange system with a Lagrangian $\mathscr L(q,\dot q)$ admits Hamiltonian regularization iff it is has no secondary constraints\footnote{For a proof of this proposition, see \cite{15}.}.


\chapter{Lagrangian Proof of No-Interaction Theorem}\label{appb}
In this Appendix, we will give an economic proof of the No-Interaction Theorem in the Lagrangian formalism. \textit{The fundamental assumption that we need to make to prove the theorem is the existence of a Lagrangian function for the system, whether it is degenerate or not}. The strategy is to let act canonically the Poincar\'e group on the Lagrangian 2-form. This will fix at each step some features of the Lagrangian. At the last step of the proof we will be able to say that the Lagrangian has a completely separated form, i.e., it can be written in the form:

\begin{equation}
\label{nit9}
\mathscr L(q,\dot q)=\sum_a\mathscr L^{(a)}(q_a,\dot q_a).
\end{equation}

\noindent
where $a$ is the particle's label. Since the Lagrangian does not mix the particle's label, each particle moves independently from the others and then there are not interactions. 

First of all, we need to introduce the fundamental quantities in the Dirac Generator Formalism. The independent coordinates of $Q$ will be written as $q_{aj}$, where $a\in(1,\dots,N)$ is the particle label and $j\in(1,2,3)$ is the Cartesian vector indices. 

Let us consider a Lagrangian function $\mathscr L=\mathscr L(q,\dot q)$ for this system of $N$ relativistic particles. From the general form of the Lagrangian $2$-form $\omega_\mathscr L$ (see Eq. (\ref{nds147})) we have the following relations:

\begin{equation}
\begin{split}
\label{nit1}
&\omega_{\mathscr L}\left(\frac{\partial}{\partial{\dot{q_{aj}}}},\frac{\partial}{\partial{\dot{q_{bk}}}}\right)=0\,\\
&\omega_{\mathscr L}\left(\frac{\partial}{\partial{\dot{q_{aj}}}},\frac{\partial}{\partial{q_{bk}}}\right)=\frac{\partial^2\mathscr L}{\partial\dot{q_{aj}}\partial{\dot{q_{bk}}}}\,\\
&\omega_{\mathscr L}\left(\frac{\partial}{\partial{q_{aj}}},\frac{\partial}{\partial{q_{bk}}}\right)=\frac{1}{2}\biggl(\frac{\partial^2{\mathscr L}}{\partial{q_{aj}}\partial{\dot{q_{bk}}}}-\frac{\partial^2{\mathscr L}}{\partial{q_{bk}}\partial{\dot{q_{aj}}}}\biggr).
\end{split}
\end{equation}

We suppose that dynamical vector field $\Gamma\in\mathfrak X(TQ)$, solution of the Euler-Lagrange equations of motion (\ref{pre8}), has the second-order form:

\begin{equation}
\begin{split}
\label{nit2}
&\Gamma=\sum_a\Gamma^{(a)}\\
&\Gamma^{(a)}=\sum_j\left(\dot q_{aj}\frac{\partial}{\partial q_{aj}}+A_{aj}\frac{\partial}{\partial\dot q_{aj}}\right),
\end{split}
\end{equation}

\noindent
where we have denoted the accelerations due to the interactions by $A_{aj}$.

We will assume that the entire Lie algebra of the Poincar\'e group $\mathcal P$ is represented by vector fields on $TQ$ which generates canonical transformations, namely:

\begin{equation}
\label{nit3}
L_{X_{P_j}}\omega_\mathscr L=L_{X_{J_j}}\omega_\mathscr L=L_{X_{K_j}}\omega_\mathscr L=L_\Gamma\omega_\mathscr L=0,
\end{equation}

\noindent
where $X_{P_j}$, $X_{J_j}$, $X_{K_j}$ are the vector fields generating spatial translations, spatial rotations and pure Lorentz transformations. Obviously, the dynamical vector field $\Gamma$ generates "time" translations and it acts canonically because it satisfies the Euler-Lagrange equations. This set of ten generators obeys to the Lie algebra of the Poincar\'e group. While the expression of the vector fields for the spatial rotations and translations are the standard ones, the form of the boosts require more attention. Indeed, from the Lie algebra of the Poincar\'e group we have that:

\begin{equation}
\label{nit4}
\left[X_{K_j},X_{P_k}\right]=\delta_{jk}\Gamma.
\end{equation}

\noindent
This means that, since on the r.h.s. there is the information on the interaction and the spatial rotations are the standard ones, the unique terms which can encode the interaction are the boosts. The explicit expression of the boosts can be obtained by implementing the WLC. In the form of the Tangent Bundle language, the WLC is the requirement that\footnote{See Sec. \ref{s232}.}:

\begin{equation}
\label{nit5}
L_{X_{K_j}}q_{ak}=q_{aj}\dot q_{ak}.
\end{equation}

\noindent
This condition fix completely the expression of the boosts vector fields:

\begin{equation}
\label{nit6}
X_{K_j}=A_j\frac{\partial}{\partial q_j}+B_j\frac{\partial}{\partial\dot q_j}.
\end{equation}

\noindent
Indeed:

\begin{equation}
L_{X_{K_j}}q_ak=i_{X_{K_j}}dq_{ak}=A=q_{aj}\dot q_{ak}
\end{equation}

\noindent
and

\begin{equation}
\begin{split}
L_{X_{K_j}}\dot q_{ak}=L_{X_{K_j}}L_\Gamma q_{ak}=L_{X_{P_j}}q_{ak}&+L_\Gamma(L_{X_{K_j}}q_{ak})=-\delta_{jk}+L_\Gamma(q_{aj}\dot q_{ak})\\
B&=-\delta_{jk}+\dot q_{aj}\dot q_{ak}+q_{aj} A_{ak},
\end{split}
\end{equation}

\noindent
where we have used the Poincar\'e Lie algebra relation $[X_{K_j},\Gamma]=X_{P_j}$. Hence the generators of the boosts becomes:

\begin{equation}
\label{nit7}
X_{K_j}=\sum_{ak}\left[q_{aj}\dot q_{ak}\frac{\partial}{\partial q_k}+(-\delta_{jk}+\dot q_{aj}\dot q_{ak}+q_{aj}A_{ak})\frac{\partial}{\partial\dot q_k}\right]
\end{equation}

\noindent
which can be written in terms of the dynamical vector field $\Gamma$ as:

\begin{equation}
\label{nit8}
X_{K_j}=\sum_a q_{aj}\Gamma^a+\sum_{ak}(\dot q_{aj}\dot q_{ak}-\delta_{jk})\frac{\partial}{\partial\dot q_k}.
\end{equation}

\noindent
As expected the generators of the boosts include the information of the interaction and then we have to use these vector fields with the conditions (\ref{nit3}) which define a canonical realization of the Poincar\'e group relative to the dynamical vector field $\Gamma$.

We are ready to prove the No-Interaction Theorem.

\begin{proof}
The proof proceeds in three steps: 

\begin{itemize}

\item[\textit{Step I)}] Apply $L_{X_{K_m}}$ to the first of Eqs. (\ref{nit1}). We get:

\begin{equation}
\label{nit10}
\omega_\mathscr L\left[\left[X_{K_m},\frac{\partial}{\partial\dot q_{aj}}\right],\frac{\partial}{\partial\dot q_{bk}}\right]=0,
\end{equation}

\noindent
where we have used Eq. (\ref{nit3}). By using the boost vector field expression (\ref{nit7}), Eq. (\ref{nit10}) becomes:

\begin{equation}
\label{nit11}
(q_{am}-q_{bm})\,\omega_{\mathscr L}\left[\frac{\partial}{\partial\dot q_{aj}},\frac{\partial}{\partial q_{bk}}\right]=(q_{am}-q_{bm})\frac{\partial^2\mathscr L}{\partial\dot q_{aj}\partial\dot q_{bk}}=0.
\end{equation}

\noindent
From Eq. (\ref{nit11}) we can conclude that for distinct particles ($a\neq b$) the second of Eqs. (\ref{nit1}) vanishes and then we can decompose tha Lagrangian w.r.t. the velocities:

\begin{equation}
\label{nit12}
\mathscr L(q,\dot q)=\sum_a\mathscr L^{(a)}(q,\dot q_a).
\end{equation}

\item[\textit{Step II})] To the result of the previous step (i.e., (\ref{nit1}.b) vanishes for different particles) we apply $L_\Gamma$. We then obtain:

\begin{equation}
\label{nit13}
\left[\frac{\partial}{\partial q_{aj}},\frac{\partial}{\partial q_{bk}}\right]=-\sum_{cl}\frac{\partial A_{al}}{\partial\dot{q_{bk}}}\omega_{\mathscr L}\left[\frac{\partial}{\partial q_{aj}},\frac{\partial}{\partial \dot q_{cl}}\right].
\end{equation}

\noindent
From the result of the Step I we know that only the term with $c=a$ survives on the r.h.s. and then Eq. (\ref{nit13}) becomes:

\begin{equation}
\label{nit14}
\left[\frac{\partial}{\partial q_{aj}},\frac{\partial}{\partial q_{bk}}\right]=-\sum_{l}\frac{\partial A_{cl}}{\partial\dot{q_{bk}}}\omega_\mathscr L\left[\frac{\partial}{\partial q_{aj}},\frac{\partial}{\partial \dot q_{al}}\right],\quad a\neq b.
\end{equation}

\noindent
Next, we apply $L_{X_{K_m}}$ on the result of the Step I and after some algebra \cite{21} we get:

\begin{equation}
\label{nit15}
q_{bm}\left[\frac{\partial}{\partial q_{aj}},\frac{\partial}{\partial q_{bk}}\right]=-q_{am}\sum_{l}\frac{\partial A_{al}}{\partial\dot{q_{bk}}}\omega_{\mathscr L}\left[\frac{\partial}{\partial q_{aj}},\frac{\partial}{\partial \dot q_{al}}\right],\quad a\neq b.
\end{equation}

\noindent
By comparing the results (\ref{nit14}) and (\ref{nit15}) we see that:

\begin{equation}
\label{nit16}
(q_{am}-q_{bm})\omega_\mathscr L\left[\frac{\partial}{\partial q_{aj}},\frac{\partial}{\partial q_{bk}}\right]=0,
\end{equation}

\noindent
from which we can conclude that:

\begin{equation}
\label{nit17}
\omega_\mathscr L\left[\frac{\partial}{\partial q_{aj}},\frac{\partial}{\partial q_{bk}}\right]=0\quad a\neq b.
\end{equation}

\noindent
From Eq. (\ref{nit17}) and the already separated form of the Lagrangian (\ref{nit12}) we can argue the following further decomposition of the Lagrangian:

\begin{equation}
\label{nit18}
\mathscr L^{(a)}(q,\dot q_a)=\mathscr L^{(a)}_{nl}(q_a,\dot q_a)-V^{(a)}(q),
\end{equation}

\noindent
where the linear terms in Eq. (\ref{nit18}) can be dropped because they amount to a total time derivative in $\mathscr L$. This equation can be understood if one considers that in any non linear dependence of the Lagrangian on the velocities, the positions cannot occour for different particles. Moreover, in any linear dipendence of the Lagrangian on the velocities, it has to satisfy Eq. (\ref{nit18}). Thus, the Lagrangian takes the completely separated form (\ref{nit18}). Altough at this stage the 2-form has achieved a completely separated form:

\begin{equation}
\label{nit19}
\begin{split}
&\omega_\mathscr L=\sum_a\omega_{\mathscr L}^{(a)},\\
&\omega_{\mathscr L}^{(a)}=\sum_j d\left(\frac{\partial \mathscr L^{(a)}_{nl}}{\partial\dot q_{aj}}\right)\wedge dq_{aj},
\end{split}
\end{equation}

\noindent
we still have the arbitrary on defining each $\mathscr L^{(a)}_{nl}$ up to a function which depends only on $q_a$. This arbitrary could have some consequences on the $V^{(a)}$ which can still generate some residual interaction.

\item[\textit{Step III})] The strategy is similar to the Step II. We have to apply $L_\Gamma$ and $L_{X_{K_m}}$ to Eq. (\ref{nit17}) and then compare the results. After some algebra \cite{21} we get the main result of the third step:

\begin{equation}
\label{nit20}
\frac{\partial}{\partial{q_{bk}}}\sum_l\omega_{\mathscr L}^{(a)}\left[\frac{\partial}{\partial q_{aj}},\frac{\partial}{\partial\dot q_{al}}\right]A_{al}=0,\quad a\neq b,
\end{equation}

\noindent
which is just equivalent to say that

\begin{equation}
\label{nit21}
\sum_l\omega_{\mathscr L}\left[\frac{\partial}{\partial q_{aj}},\frac{\partial}{\partial\dot q_{al}}\right]A_{al}
\end{equation}

\noindent
can depend on $q_a$, $\dot q_a$ and possibly on $q_b$ for $b\neq a$ via $A_{al}$. If we write Eq. (\ref{nit20}) in local coordinates we get:

\begin{equation}
\label{nit22}
-\frac{\partial\mathscr{L}^{(a)}_{nl}}{\partial q_{aj}}(q_a,\dot q_a)+\sum_k\left(\frac{\partial^2\mathscr{L}^{(a)}_{nl}}{\partial\dot q_{aj}\partial\dot q_{ak}}A_{ak}+\frac{\partial^2\mathscr{L}^{(a)}_{nl}}{\partial\dot q_{aj}\partial q_ak}\dot q_{ak}\right)=-\frac{\partial V(q)}{\partial_{aj}},
\end{equation}

\noindent
where $V(q)=\sum_a V^{(a)}(q)$. Now, since the l.h.s. has no dependence on $q_b$ for $b\neq a$ we can conclude that the "potential" terms is completely separable, i.e.:

\begin{equation}
\label{nit23}
V(q)=\sum_a V^{(a)}(q_a)
\end{equation}

\noindent
and we have arrived at a completely separated form for the Lagrangian function:

\begin{equation}
\label{nit24}
\mathscr L(q,\dot q)=\sum_a \mathscr{L}^{(a)}_{nl}(q_a,\dot q_a)-V^{(a)}(q_a)
\end{equation}

\noindent
from which we conclude that there are no interactions.

\end{itemize}
\end{proof}


\chapter{Computation of the Dirac Brackets}\label{appdb}
Let us explicitly construct the DB for the set of constraints:

\begin{equation}
\label{constraints}
\begin{split}
&K_1=p^2_1-m_1^2+V(\xi),\qquad\chi_1(x,p)=P\cdot r\\
&K_1=p^2_2-m_2^2+V(\xi),\qquad\chi_2(x,p,\tau)=P\cdot\frac{(x_1+x_2)}{2}-\tau
\end{split}
\end{equation}

\noindent
where $P=p_1+p_2$, $r=(x_1-x_2)/2$, $\xi=r^2-(P\cdot r)^2/p^2$. We have to build the matrix of the constraints (\ref{anit19}). We start from:

\begin{equation}
\label{anit21}
\begin{split}
\{\chi_1,K_1\}&=\{(P\cdot r),p_1^2\}+\{(P\cdot r),\xi\}V'\\
&=2\{(P\cdot r),p_1\}p_1+\left(\{P\cdot r,r^2\}-\left\{P\cdot r,\frac{(P\cdot r)^2}{p^2}\right\}\right)V'\\
&=(p_1+p_2)p_1+\left[\{P\cdot r,x_1\}-\{(P\cdot r,x_2\}r\right]V'-\{P\cdot r,\frac{(P\cdot r)^2}{p^2}\}V'\\
&=p_1^2+p_2p_1+V'
\end{split}
\end{equation}

\begin{equation}
\label{anit22}
\begin{split}
\{\chi_2,K_1\}&=\frac{1}{2}\biggl[2\{P\cdot(x_1+x_2),p_1\}p_1+\{P\cdot(x_1+x_2),\xi\}V'\biggr]\\
&=\frac{1}{2}\left[2(p_1+p_2)p_1+(\{P\cdot(x_1+x_2),r^2\}-\{P\cdot(x_1+x_2),\frac{(P\cdot r)^2}{p^2}\})V'\right]\\
&=\frac{1}{2}\biggl[2(p_1^2+p_2p_1)-\{P\cdot(x_1+x_2),P\cdot(x_1-x_2)\}\frac{1}{p^2}+2\{P\cdot(x_1+x_2),p\}\frac{(P\cdot r)^2}{p^4}\biggr]\\
&=p_1^2+p_1p_2+\frac{(P\cdot r)^2}{p^4}
\end{split}
\end{equation}

\begin{equation}
\label{anit23}
\begin{split}
\{\chi_1,K_2\}&=\{P\cdot r,p_2^2\}+\{P\cdot r,\xi\}V'\\
&=\{P\cdot r,p_2\}p_2+\frac{2}{p^4}\{P\cdot r,p\}V'\\
&=p_1p_2+p_2^2+\frac{2}{\pi^4}V'
\end{split}
\end{equation}

\begin{equation}
\label{anit24}
\begin{split}
\{\chi_2,K_2\}&=\frac{1}{2}\biggl[\{P\cdot(x_1+x_2),p_2^2\}+\{P\cdot(x_1+x_2),\xi\}V'\biggr]\\
&=\frac{1}{2}\left[2(p_1p_2+P_2^2)+2\{P\cdot(x_1+x_2),\pi\}\frac{(P\cdot r)^2}{p^4}\right]\\
&=p_1p_2+p_2^2+\frac{(P\cdot r)^2}{p^4}
\end{split}
\end{equation}

\noindent
Then, the constraint matrix $\mathcal A$ is 

\begin{equation}
\label{anit25}
\mathcal A=\begin{pmatrix}
p_1^2+p_2p_1+V'&p_1p_2+p_2^2+\frac{2}{\pi^4} V'\\
p_1^2+p_1p_2+\frac{(P\cdot r)^2}{p^4}&p_1p_2+p_2^2+\frac{(P\cdot r)^2}{p^4}
\end{pmatrix}
\end{equation}

\noindent
Its determinant is:

\begin{equation}
\label{anit26}
det\,\mathcal A=(p_1^2-p_2^2)\left(\frac{(P\cdot r)^2}{p^4}-\frac{2}{p^4}V'\right)
\end{equation}

\noindent
and then the inverse matrix $\mathcal A^{-1}$ is:

\begin{equation}
\label{anit27}
\mathcal A^{-1}=\begin{pmatrix}
p_2^2+p_1p_2+\frac{(P\cdot r)^2}{p^4}&-p_2-p_1p_2-\frac{2}{p^4}V'\\
-p_1^2-p_1p_2-\frac{(P\cdot r)^2}{p^4}&p_1^2+p_1p_2+\frac{2}{p^4}V'
\end{pmatrix}\frac{1}{det\,\mathcal A}
\end{equation}

\noindent
Finally, we can define the DB:

\begin{equation}
\label{anit28}
\{f,g\}_D=\{f,g\}-\sum_{\alpha\beta}\{f,v_\alpha\}A^{-1}_{\alpha\beta}\{v_\beta,g\}
\end{equation}

\noindent
where the $v$'s runs on the constraints set.


\chapter{From Commutation Relations to Lagrangian Function}\label{appf}
In this Appendix we will show that if the commutation relations:

\begin{equation}
\label{anit32}
i)\,\{q^i,q^j\}=0,\quad\ ii)\,\{q^i,\dot q^j\}=g^{ij},\quad iii)\,\{\dot q^i,\dot q^j\}=f^{ij}(q,\dot q),
\end{equation}

\noindent
hold, where $f^{ij}$ represents the component of a skew-symmetric tensor with the property that it satisfies the Jacobi identity, then a Lagrangian function must exists\footnote{The problem of describing the evolution of a dynamical systems by giving just the right commutation relations w.r.t. the PB, without defining a Lagrangian function, was aimed by Richard Feynman for the first time (see \cite{4} for details).}.
\\

In Appendix \ref{s21} we show that, also on the Tangent Bundle, it is possible to write the PB in terms of a Poisson tensor:

\begin{equation}
\label{anit33}
\{f,g\}=\Lambda(df,dg),
\end{equation}

\noindent
Conditions i) and ii) imply that:

\begin{equation}
\label{anit34}
\Lambda(dq^i,\cdot)=\frac{\partial}{\partial\dot q^j}.
\end{equation}

\noindent
Since the Poisson tensor is defined as the "inverse" of a non-degenerate Lagrangian 2-form $\omega$, i.e.: $\omega(\Lambda)=\mathds 1$, we can write Eq. (\ref{anit34}) in a covariant form:

\begin{equation}
\label{anit35}
\omega\left(\frac{\partial}{\partial\dot q^j},\cdot\right)=dq^i.
\end{equation}

\noindent
This means that vertical bivector field $\frac{\partial}{\partial\dot q^i}\wedge\frac{\partial}{\partial\dot q^j}$ is in the kernel of the 2-form $\omega$ under contraction. Moreover, the vector field:

\begin{equation}
\label{anit36}
\Gamma=g^{ij}\frac{\partial}{\partial q^i}+f^{ij}\frac{\partial}{\partial \dot q^i}
\end{equation}

\noindent
satisfies the condition $L_\Gamma\omega=0$. Therefore, the $2$-form $\omega$ satisfies all the requirement of Prop. (\ref{p21}) and then it can be written in terms of a function $\mathscr L$ that we identify as the Lagrangian function, as:

\begin{equation}
\label{anit37}
\omega=\frac{\partial^2{\mathscr{L}}}{\partial{v^i}\partial{v^j}}dq^i\wedge dv^j+\frac{1}{2}\left(\frac{\partial^2{\mathscr{L}}}{\partial{v^i}\partial{q^j}}-\frac{\partial^2{\mathscr{L}}}{\partial{v^j}\partial{q^i}}\right)dq^i\wedge dq^j.
\end{equation}


\chapter*{Ringraziamenti}
La tesi magistrale, secondo l'opinione di molti, non dovrebbe contenere un capitolo di ringraziamenti. Tuttavia, chi mi conosce sa bene quanto conti per me l'opinione dei molti.

Desidero innanzitutto ringraziare il professor Giuseppe Marmo per essere stato il miglior relatore che uno studente possa desiderare. Nella sua \textit{bottega} ho potuto vivere un'esperienza intensa e formativa, entrando in contatto con persone straordinarie e luoghi meravigliosi. Ho scoperto i miei limiti e, con il suo aiuto, sto imparando a superarli.

Vorrei ringraziare la professoressa Vitale per aver tenuto sempre la sua porta aperta e per aver trovato sempre il tempo per ascoltarmi ed il professor Lizzi per le utilissime discussioni sul capitolo conclusivo della tesi.

Tra le varie cose che chi mi conosce ben sa, c'\`e il mio distaccato rapporto con la religione. Questa mia convinzione \`e stata fortemente messa in crisi dopo aver conosciuto Fabio di Cosmo e Florio Maria Ciaglia. Questi due signori, non solo sono stati dei riferimenti accademici disponibili e straordinari, non solo sono stati compagni di viaggio e di avventure, ma sono stati per me, e spero lo saranno ancora dopo questa dimostrazione di affetto, dei veri \textit{amici}. Grazie dal profondo del cuore.

Vorrei ringraziare Elisabetta, a cui questa tesi \`e giustamente dedicata. Senza di lei non avrei mai avuto il coraggio di iniziare questa tesi. Devo a lei anche la mia decisione di concentrarmi sulla Fisica Teorica. Elisabetta \`e stata il carburante che mancava a tutta la mia vita. Posso dire, con dignit\`a e cognizione di causa, che senza di lei non sarei arrivato fin qui. Grazie.

Un prodotto del mio percorso magistrale di cui sono molto felice \`e la mia amicizia con Fabio Mele. Con questa persona straordinaria ho affrontato una delle sfide accademiche pi\`u difficili che mi si sono presentate. Non dimenticher\`o mai le appassionanti discussioni a casa sua e le nottate infinite per completare la stesura delle note. Spero con tutto il cuore che la nostra collaborazione possa continuare anche in futuro.

Desidero ringraziare tutte le persone che mi hanno accompagnato durante il mio percorso accademico, in particolare Giorgio, Luca, Daniele, Anna, Pierpaolo e Michele, che sono stati con me sempre, fin dall'inizio. Vorrei ringraziare il mitico gruppo degli astrofisici, in particolare Giulia per essere stata per me un costante esempio di notevole forza d'animo. Vorrei ringraziare Luca Buonocore per quel bellissimo pomeriggio dedicato alla Relativit\`a Generale.

Vorrei infine ringraziare la mia famiglia. Innanzitutto mia madre e mio padre. Inutile dire che senza il loro supporto nessuna di queste righe sarebbe mai stata scritta. Questo lavoro \`e dedicato anche a voi. Vorrei in particolare ringraziare mio fratello Diego. Ripensando al mio percorso magistrale mi sono accorto di come ogni volta che volevo parlare dei miei problemi andavo instintivamente nella stanza di fronte a parlare con lui. Ti auguro un grande in bocca al lupo per le tue (scellerate) scelte accademiche. Voglio ringraziare anche i miei nonni per l'infinito affetto che mi dimostrano ogni volta.

Vorrei ringraziare anche la mia seconda famiglia composta da Genny, Rosy, Giovanni, Ludovica e Nunzia per avermi accolto e per avermi fatto sentire sempre come una parte della famiglia.

Vorrei concludere questa stucchevole serie di ringraziamenti con l'augurio di poter ricordare tutte queste persone nei miei futuri lavori.
\end{document}